\documentclass[3p,onecolumn]{elsarticle}

\usepackage{amssymb}
\usepackage{lineno,hyperref}
\usepackage{hyperref}
\usepackage{color}
\modulolinenumbers[5]

\journal{ }

\begin{document}

\begin{frontmatter}

\title{CAPP-8TB: Axion Dark Matter Search Experiment around 6.7\,$\mu$eV}

\author[CAPP]{J. Choi\fnref{fn1}}
\author[KAIST]{S. Ahn}
\author[CAPP]{B. R. Ko}
\author[CAPP]{S. Lee\corref{cor}}
\ead{soohyunglee@ibs.re.kr}
\author[CAPP,KAIST]{Y. K. Semertzidis}

\cortext[cor]{Corresponding author}
\fntext[fn1]{Now at Korea Astronomy and Space Science Institute, Daejeon 34055, Republic of Korea}

\address[CAPP]{Center for Axion and Precision Physics Research, Institute for Basic Science, Creation Hall, 193 Munji-ro, Yuseong-gu, Daejeon, Republic of Korea, 34051}
\address[KAIST]{Korea Advanced Institute of Science and Technology, 291, Daehak-ro, Yuseong-gu, Daejeon, Republic of Korea, 34141}

\begin{abstract}
{CAPP-8TB is an axion dark matter search experiment dedicated to an axion mass search near 6.7\,$\mu$eV. The experiment uses a microwave resonant cavity under a strong magnetic field of 8\,T produced by a superconducting solenoid magnet in a dilution refrigerator. We describe the experimental configuration used to search for a mass range of 6.62 to 6.82\,$\mu$eV in the first phase of the experiment. We also discuss the next phase of the experiment and its prospects.}
\end{abstract}

\begin{keyword}
Axion, Dark matter, Axion haloscope, Microwave cavity, Superconducting magnet
\end{keyword}

\end{frontmatter}


\section{Introduction}
\label{sect:introduction}
The strong $CP$ problem remains one of the biggest mysteries in particle physics. A theoretical solution was proposed by Peccei and Quinn introduced the Peccei-Quinn ($PQ$) symmetry \cite{PhysRevLett_38_1440_1977}, and Weinberg \cite{PhysRevLett_40_223_1978} and Wilczek \cite{PhysRevLett_40_279_1978} found that the $PQ$ symmetry results in a pseudo Goldstone boson, designated as the axion. By nature, the axion is also considered to be a promising candidate for dark matter \cite{NewJPhys_11_105008_2009}. 

The original axion proposed by Weinberg and Wilczek, which is known as the PQWW axion, was quickly ruled out by experimental results \cite{PhysRep150_1_1987}. Two other models have been proposed, the Kim-Shifman-Vainshtein-Zakharov (KSVZ) \cite{PhysRevLett43_103_1979,NuclPhysB166_493_1980} and Dine-Fischler-Srednicki-Zhitnitsky (DFSZ) \cite{PhysLettB104_199_1981,SovJNuclPhys31_260_1980}.

Over decades, many experiments have attempted to discover axion dark matter using an experimental technique proposed by Sikivie \cite{PhysRevLett_51_1415_1983}. Based on the expectation that the axion converts into a photon under a strong magnetic field, a microwave resonator is used as a detector to observe the electromagnetic wave produced by the axion-photon conversion. The experimental searches are focused in a mass range from 1\,$\mu$eV \cite{PhysLett_120B_127_1983,PhysLett_120B_133_1983,PhysLett_120B_137_1983} to 3\,meV \cite{PhysLettB_193_525_1987,PhysRevLett_60_1793_1988,PhysRevLett_60_1797_1988,PhysRevLett_76_2621_1996,PhysRevD_56_2419_1997}.

\begin{figure*}[t]
\centerline{\includegraphics[width=.95\textwidth]{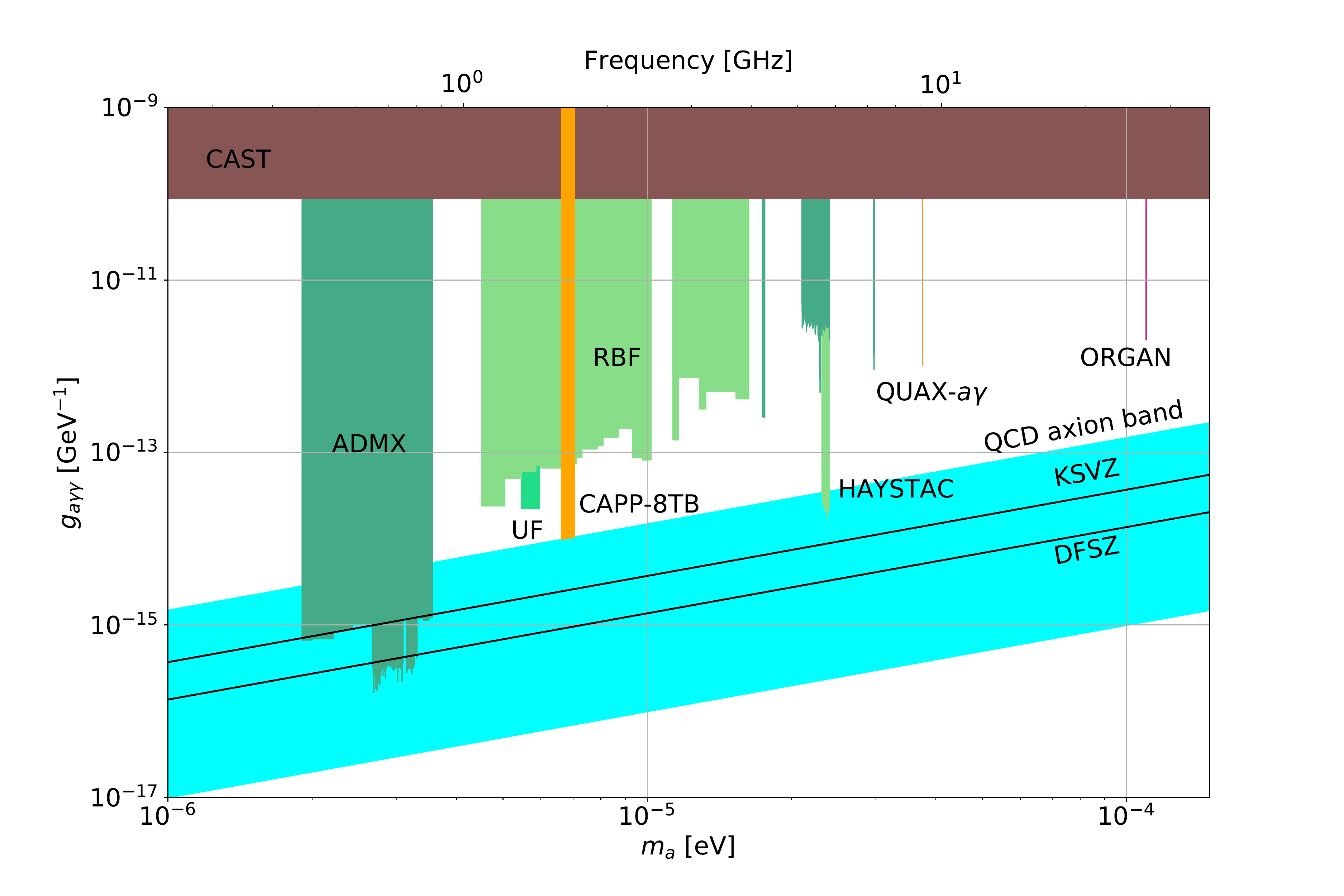}}
\caption{Excluded axion mass and axion-photon coupling for axion haloscope experiments by Rochester-Brookhaven-FNAL (RBF) \cite{PhysRevD40_3153_1989}, University of Florida (UF) \cite{PhysRevD42_1297_1990}, ADMX \cite{PhysRevLett80_2043_1998,AstroJLett571_L27_2002,PhysRevD69_011101_2004,PhysRevLett104_041301_2010,PhysRevLett120_151301_2018,PhysRevLett124_101303_2020,PhysRevLett121_261302_2018}, HAYSTAC \cite{PhysRevLett118_061302_2017,PhysRevD97_092001_2018}, ORGAN \cite{PhysDarkUniverse_18_67_2017}, and QUAX-$a\gamma$ \cite{PhysRevD99_101101_2019}, and a project of the CAPP-8TB experiment. The result from CAST \cite{NaturePhys_13_584_2017} is also shown. Theoretical expectations of the KSVZ \cite{PhysRevLett43_103_1979,NuclPhysB166_493_1980}, DFSZ \cite{PhysLettB104_199_1981,SovJNuclPhys31_260_1980}, and QCD axion model band \cite{PhysRevD52_3132_1995} are shown together as well.}\label{fig:sensitivity}
\end{figure*}

Over this mass region, CAST set the upper limit of $0.66\times10^{-10}$\,GeV$^{-1}$ \cite{NaturePhys_13_584_2017}. Rochester-BNL-FNAL (RBF) excluded a broad range of 4.5 to 16.3\,$\mu$eV with a coupling of about $10^{-13}$ to $10^{-12}$\,GeV$^{-1}$ \cite{PhysRevD40_3153_1989}. University of Florida (UF) excluded 5.4 to 5.9\,$\mu$eV with a coupling near $10^{-14}$\,GeV$^{-1}$ \cite{PhysRevD42_1297_1990}. ADMX \cite{PhysRevD64_092003_2001} searched a broad range of 1.9 to 3.53\,$\mu$eV with the sensitivity of the KSVZ model \cite{PhysRevLett80_2043_1998,AstroJLett571_L27_2002,PhysRevD69_011101_2004,PhysRevLett104_041301_2010}. They excluded further axion-photon coupling below the DFSZ model in a mass range of 2.66 to 3.31\,$\mu$eV \cite{PhysRevLett120_151301_2018,PhysRevLett124_101303_2020}. ADMX also scanned a mass range of 17.38 to 17.57\,$\mu$eV, 21.03 to 23.98\,$\mu$eV, and 29.67 to 29.79\,$\mu$eV with sensitivities of around $10^{-13}$ to $10^{-12}$\,GeV$^{-1}$ \cite{PhysRevLett121_261302_2018} using a parasitic microwave resonant cavity. HAYSTAC \cite{NuclInstrumMethA854_11_2017} excluded 23.15 to 24\,$\mu$eV with sensitivities near the KSVZ model \cite{PhysRevLett118_061302_2017,PhysRevD97_092001_2018}. ORGAN demonstrated a search capacity in a higher axion mass region of 109.835 to 109.840\,$\mu$eV with about $10^{-12}$\,GeV$^{-1}$ \cite{PhysDarkUniverse_18_67_2017} using a small microwave cavities, and QUAX-$a\gamma$ demonstrated the feasibility of using a superconducting cavity for axion searches by excluding a mass region of 37.499 to 37.501\,$\mu$eV with a coupling of about $10^{-12}$\,GeV$^{-1}$ \cite{PhysRevD99_101101_2019}. The current exclusions are summarized in Figure \ref{fig:sensitivity}.

\begin{figure}[t]
\centerline{\includegraphics[width=.95\textwidth]{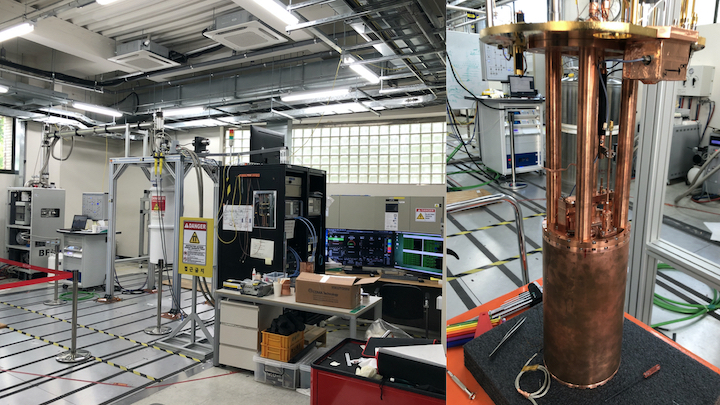}}
\caption{The CAPP-8TB experiment. Left: The experiment area. Right: The resonant cavity.}\label{fig:apparatus}
\end{figure}

The CAPP-8TB experiment is an axion haloscope search for an axion mass near 6.7\,$\mu$eV. The experiment uses a microwave resonant cavity as a detector of the axion signal under a strong magnetic field in a cryogenic environment. The apparatus of the experiment is shown in Figure \ref{fig:apparatus}. The detected power from the axion to photon conversion in a microwave resonant cavity is described with the SI unit as
\begin{eqnarray}
P_{S} & = & g^{2}_{a\gamma\gamma}\frac{\rho_{a}\hbar^{2}}{m^{2}_{a}c}\omega(2U_{M})C_{nlm}Q_{L}\frac{\beta}{1+\beta}
\label{eq:conversion_power}
\end{eqnarray}
with $g_{a\gamma\gamma}$ the axion-photon coupling strength ($g_{a\gamma\gamma}=\alpha g_{\gamma}/(\pi f_{a})$ where $\alpha$ is the fine structure constant, $g_{\gamma}$ is a model-dependent axion-photon coupling constant, and $f_{a}$ is the Peccei-Quinn scale), $\rho_{a}$ is the local dark matter density, $\hbar$ is the normalized Planck constant,  $c$ is the speed of light, $m_{a}$ is the mass of the axion ($m_{a}=h\nu/c^{2}$ where $\nu$ is the resonant frequency of the resonant cavity matched with the axion mass), $\omega=2\pi\nu$,  $U_{M}=\frac{1}{2\mu_{0}}B^{2}_{\mathrm{avg}}V$ is the energy stored in an averaged magnetic field $B_{\mathrm{avg}}$ over the cavity volume $V$, and $Q_{L}$ ($Q_{L}\ll Q_{a}$ where $Q_{a}$ is the axion quality factor), $C_{nlm}$, and $\beta$ are the loaded quality factor, the form factor, and the antenna coupling coefficient of the TM$_{nlm}$ mode of the resonant cavity, respectively.

The scan rate of the experiment is proportional to the axion conversion power squared (Eq. (\ref{eq:conversion_power})) and experimental parameters \cite{JCosmolAstropartP_03_066_2020}
\begin{eqnarray}
\frac{d\nu}{dt} & \propto & \eta\frac{B_{\mathrm{avg}}^{4}V^{2}C_{nlm}^{2}Q_{L}}{(\mathrm{SNR})^{2}T_{n}^{2}}
\label{eq:scan_rate}
\end{eqnarray}
where $\eta$ is the efficiency of the data acquisition, SNR is the signal-to-noise ratio, and $T_{n}$ is the system noise temperature.

Throughout this paper, we use the local density of dark matter, $\rho_{a}$, as 0.45\,GeV/cm$^{3}$, and the model-dependent axion-photon coupling constants, $g_{\gamma}$, as 0.97 and $-0.36$ for the KSVZ ($g_{\gamma}^{\mathrm{KSVZ}}$) and DFSZ ($g_{\gamma}^{\mathrm{DFSZ}}$) axions, respectively. To maximize the form factor, the experiment employs a TM$_{010}$ mode.

\section{Dilution Refrigerator}
\label{sect:dilution_refrigerator}
To keep the physical temperature of the system as low as possible, and thus minimize the system noise temperature in Eq. (\ref{eq:scan_rate}), a dilution refrigerator is used in the experiment. We employed the BlueFors LD400 dilution refrigerator \cite{BlueFors} for the experiment. The benefit of this cryogen-free system is its easy operation, however, it relies on a stable supply of electricity. 

To be robust against failure in continuous power supply, an uninterruptible power supply (UPS) is employed. The UPS provides power from its battery to the refrigerator to prevent any power outage. 

\begin{figure}[h]
\centerline{\includegraphics[width=.35\textwidth]{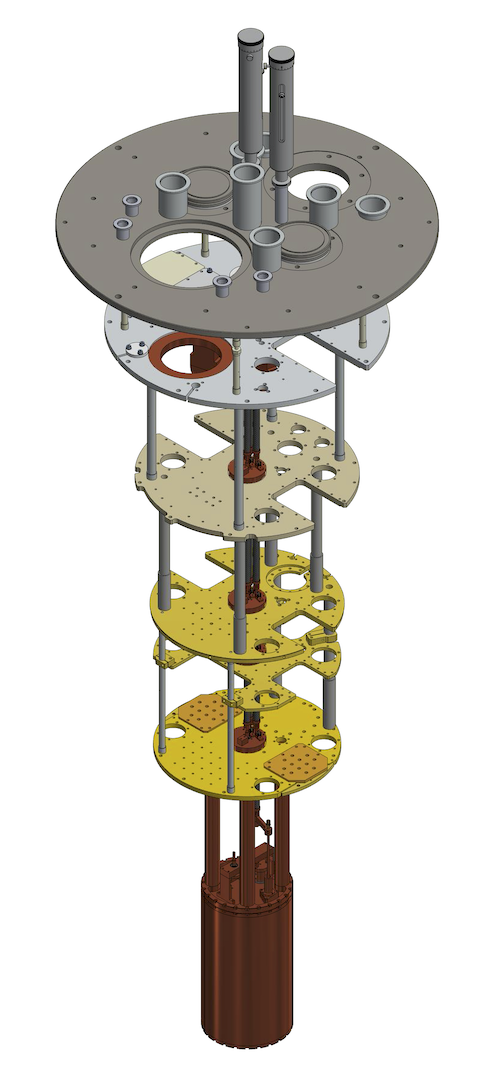}}
\caption{BlueFors dilution refrigerator, LD400, with the resonant cavity and tuning system described in later sections. From the top, there are the 300\,K, 50\,K, 4\,K, 1\,K, 100\,mK, and base temperature stages. The dilution unit of the refrigerator is not shown in this figure.}\label{fig:refrigerator}
\end{figure}

The refrigerator consists of different temperature stages of 50\,K, 1\,K, 100\,mK, and 10\,mK (10\,mK is referred to as the ``base temperature stage''). Each stage has a gold-coated plate with screw holes where various components necessary to the experiment can be mounted. Plates also have several coaxial radio-frequency (RF) ports to transmit RF signals from one to another with proper thermal links.

The physical temperature of the base temperature stage can be controlled by a heater placed on the plate. We set the physical temperature to maintain the resonant cavity at 50\,mK. 

Twelve temperature sensors made by Lake Shore Cryotronics \cite{LakeShore} were used to monitor the physical temperatures inside the system. Five sensors were placed on the 50\,K, 4\,K, 1\,K, 100\,mK, and base temperature stages to read the temperatures of each temperature stage. The temperature of the superconducting magnet was also measured. In addition, we measured the temperatures of the microwave components and resonant cavity. A Cernox sensor calibrated to the temperature range of 0.3 to 100.0\,K was used to measure the temperature of the cryogenic amplifier at the 4\,K stage; a Cernox sensor calibrated for 0.1 to 40.0\,K was used for a cryogenic amplifier at the 1\,K stage; three RuO$_{2}$ sensors calibrated for 0.05 to 40.00\,K were used for the cavity top and bottom, and an RF circulator at the base temperature stage; a RuO$_{2}$ sensor calibrated for 0.02 to 40.00\,K was used for the locomotive tuning structure described in Sect. \ref{sect:tuning}. The uncertainties of the temperature measurements did not exceed 1\,\% \cite{LakeShore}.

The refrigerator was controlled and monitored by a LabVIEW program \cite{LabVIEW} provided by the manufacturer on a control computer. The cooling power at 20\,mK was 16\,$\mu$W, and it took about 22 hours to cool the system down to the base temperature without any load inside the refrigerator.

\section{Superconducting Magnet}
\label{sect:magnet}
A superconducting magnet was installed about 26\,cm below the base temperature stage to provide the cavity with an external magnetic field, and thermally attached to the 4\,K stage. The magnet made by American Magnetics \cite{AMI} was a cryogen-free solenoid with a compensated region. The nominal magnetic field was 8\,T at an operating current of 96.56\,A, and the inductance was 53\,H. The bore and height of the magnet were 165.4 and 475.8\,mm, respectively. 

\begin{figure}[t]
\centerline{\includegraphics[width=.45\textwidth]{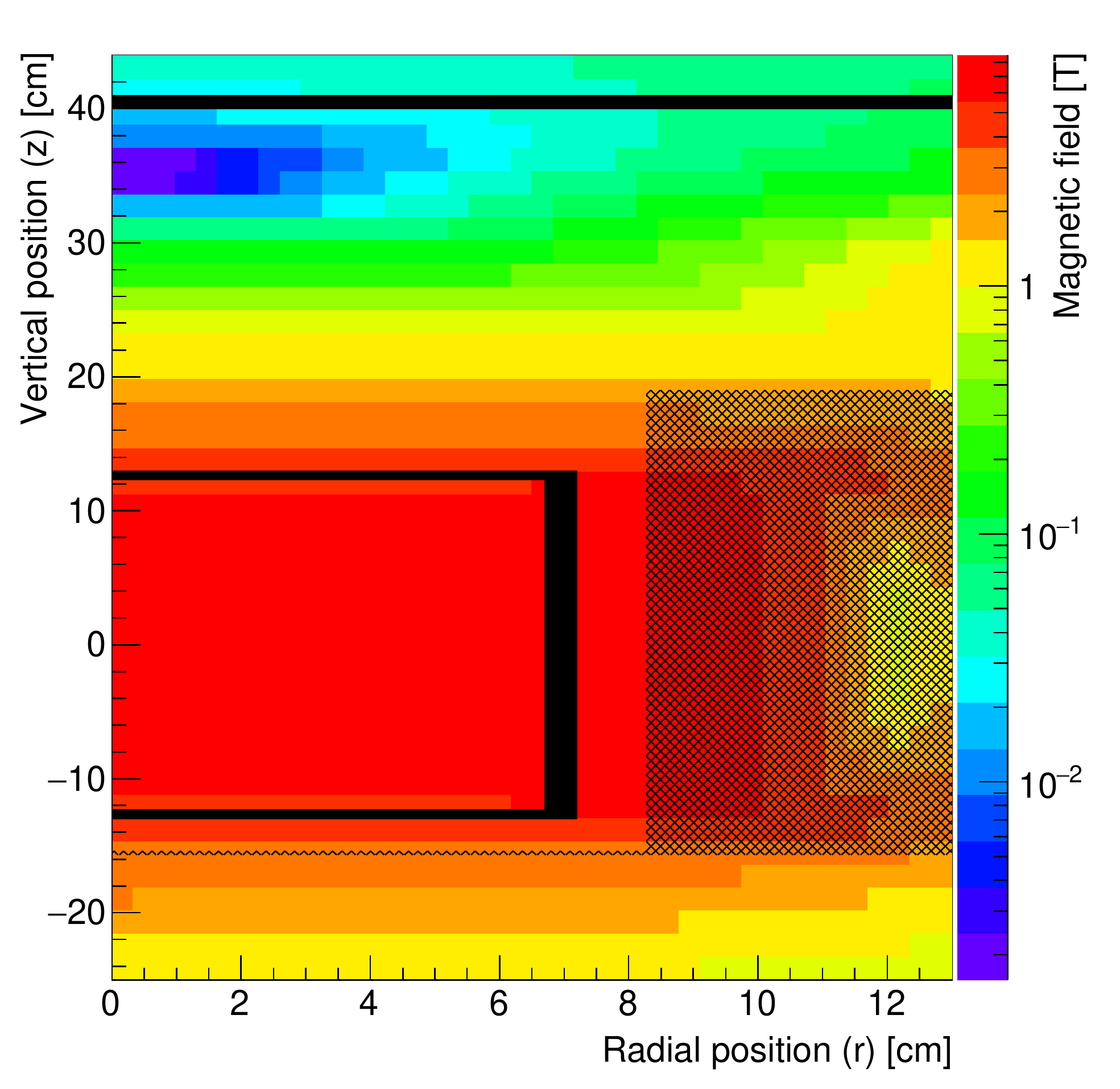}}
\caption{Magnetic field inside the superconducting magnet. The resonant cavity of the experiment is shown as a black rectangle (near $|z|=12.3$\,cm and $r=6.7$\,cm where $z$ and $r$ are vertical and radial positions, respectively), the magnet is shown as a black hatched region, and the base temperature stage is shown as a black line at $z=40$\,cm. The compensation coil of the magnet maintains a magnetic field of a few hundreds Gauss near the base temperature stage.}
\label{fig:magnetic_field_map}
\end{figure}

Figure \ref{fig:magnetic_field_map} shows the magnetic field inside the magnet volume. Since the magnet has a cancellation coil to cancel the field, the magnetic field is reduced to a few hundreds Gauss near the base temperature stage, and the RF components located at the stage function well without being affected by the strong magnetic field.

\begin{figure}[t]
\centerline{\includegraphics[width=.45\textwidth]{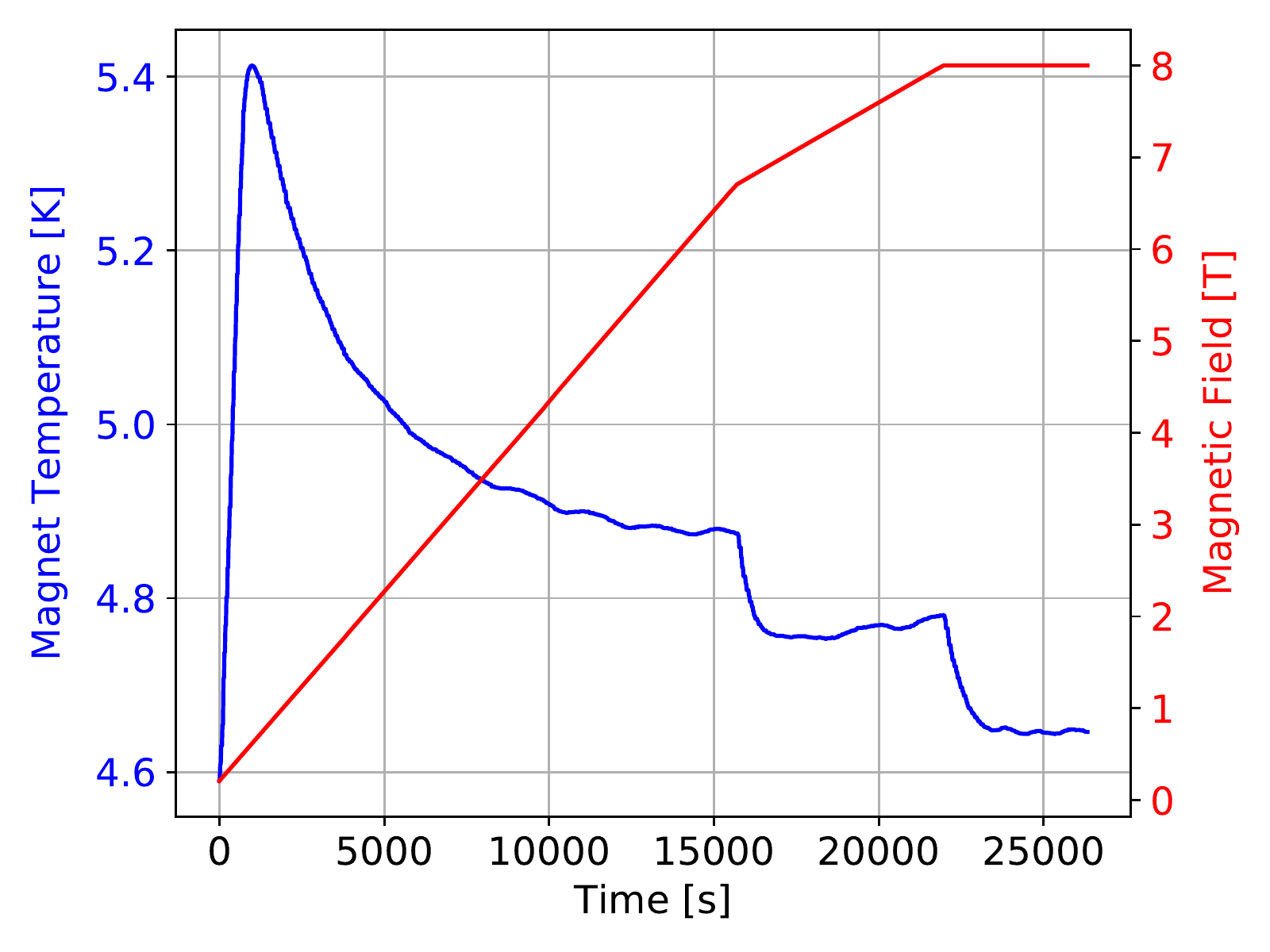}}
\caption{Magnetic field (red) and the temperature of the magnet (blue) during ramping up. Eddy currents heat up the magnet as the magnetic field changes. Once the magnet is fully charged, the system temperature goes down and is saturated.}\label{fig:magnet_temperature}
\end{figure}

For safe ramping, the magnet is ramped up and down through two segments of different ramping rates: $\pm$0.0050\,A/s for 0 to 6.7\,T and $\pm$0.0025\,A/s for 6.7 to 8.0\,T. Ramping takes about six hours, and the temperature of the magnet changes depending on the ramping rate, as shown in Figure \ref{fig:magnet_temperature}.

Like the dilution refrigerator, a stable power supply is critical for the magnet. The power supply for the magnet maintains the current within $\pm$0.15\,mA. Any power outage causes a magnet quench, and its damage is severe. To protect the system from a magnet quench due to a power outage, we employ an extra UPS of 2700\,W, and it is dedicated to the magnet and data acquisition computer. 

The UPS provides power from its battery for up to roughly 50 minutes when the magnet is in operation enough to safely discharge the magnet.

\section{Microwave Resonant Cavity}
\label{sect:cavity}
The dimensions of the magnet clear bore limits the dimensions of the resonant cavity in the experiment. Considering the volume of the magnet clear bore and the average magnetic field, we determined the inner diameter and height of the cavity to be 134\,mm and 246\,mm, respectively, therefore, the inner volume of the cavity is about 3.47 liters. Based on the magnetic field map, the average magnetic field inside the cavity volume is about 7.3\,T.

\begin{figure}[t]
\centerline{\includegraphics[width=.4\textwidth]{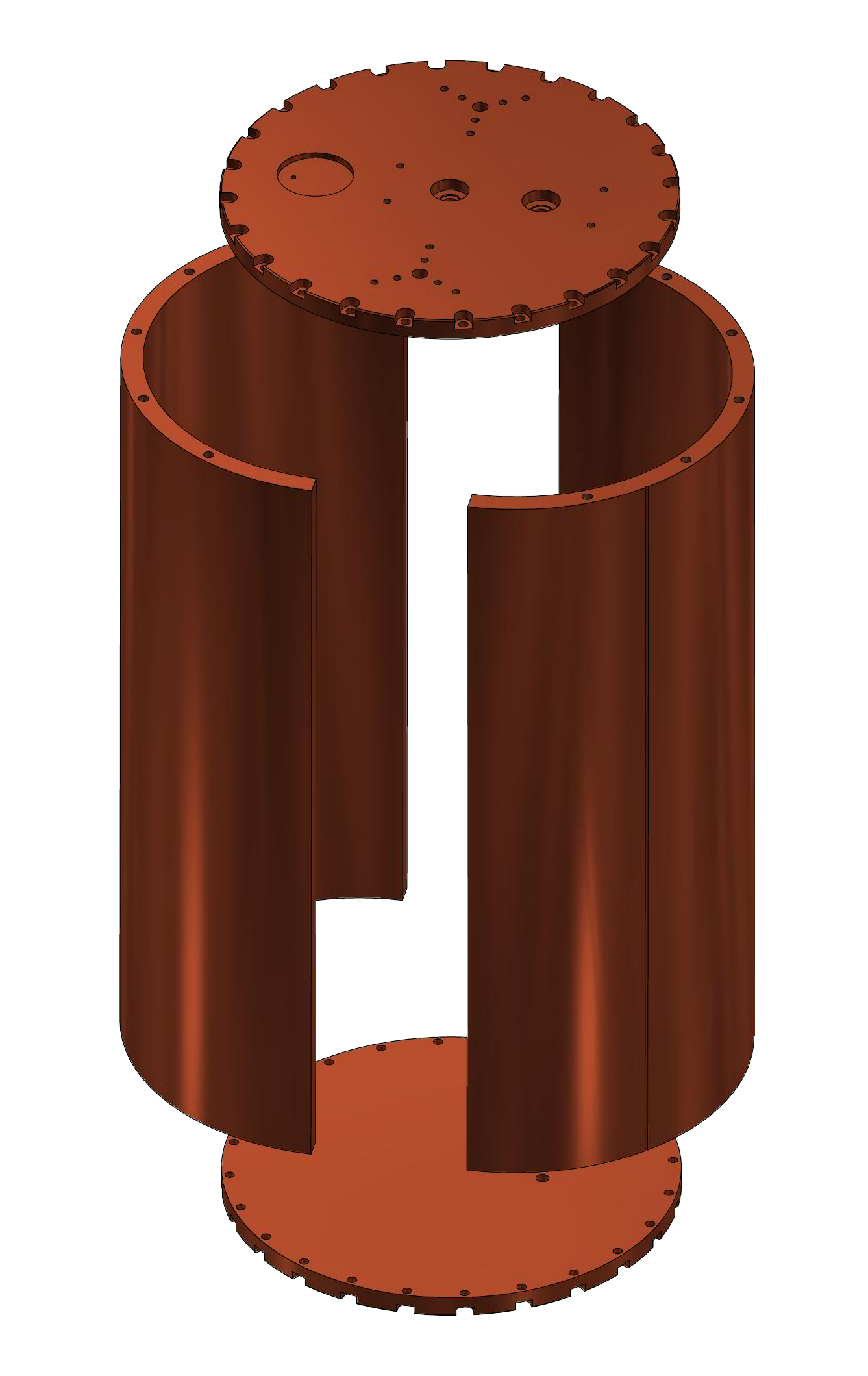}}
\caption{Microwave resonant cavity used for the experiment. To minimize Eddy currents along the cavity wall in the azimuthal direction, the barrel part of the cavity was split into two pieces. The thickness of the cavity wall is 5\,mm, and the walls are split by 0.5\,mm gap.}\label{fig:split_cavity}
\end{figure}

\begin{figure}[t]
\centerline{\includegraphics[width=.45\textwidth]{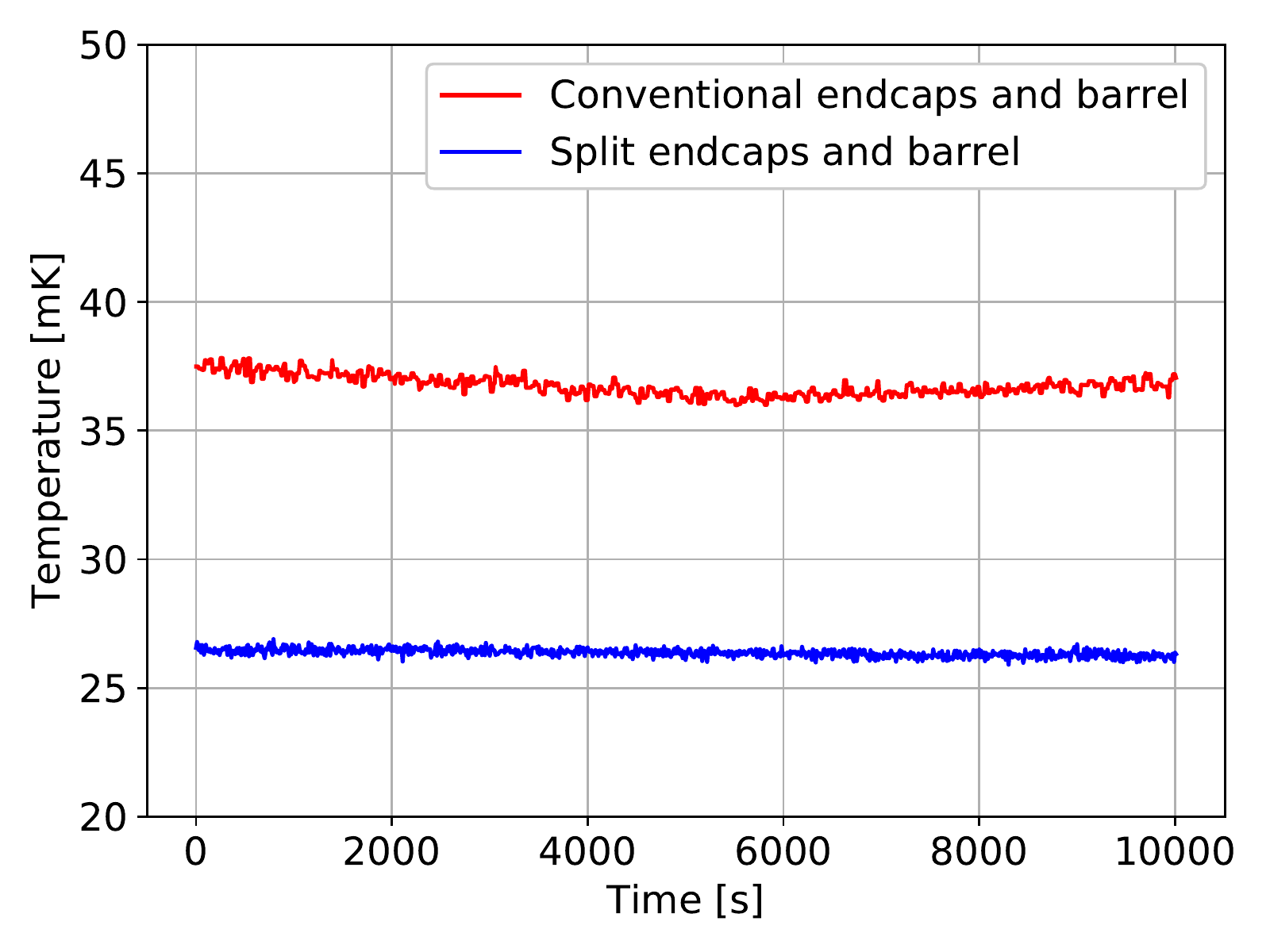}}
\caption{Physical temperatures of cavities with conventional endcaps and barrel (red), and split endcaps and barrel (blue) at the base temperature of the refrigerator under a magnetic field of 8\,T.}\label{fig:cavity_temperature}
\end{figure}

The microwave resonant cavity and related fixtures are made of an oxygen-free high conductivity copper. Since the cavity undergoes a strong magnetic field, any vibration may cause heat from Eddy currents. To minimize vibration in the system, we limited the number of physical and electrical contacts between the refrigerator and the floor in the experimental hall. In addition, we employed a split cavity, so that the barrel part of the cavity was divided into two pieces as shown in Figure \ref{fig:split_cavity}, to minimize Eddy currents. The effectiveness of the split cavity is confirmed by comparing the physical temperatures of a non-split and split cavities in the refrigerator as shown in Figure \ref{fig:cavity_temperature}.

The cavity is hung on the base temperature stage by four thick, round-shaped copper poles. There is a ring-shaped copper fixture between the poles and the cavity to connect them, and to maximize the interface for a good thermal contact. For the same reason, a copper fixture is located between the poles and the base temperature stage. The top plate and barrel of the cavity are secured by 12 screws, and an additional 12 screws connect the fixture and the barrel through the top of the cavity, to make the top and barrel connection even tighter. At the bottom of the cavity, 24 screws assemble the bottom plate and barrel.

The signal of the TM$_{010}$ mode inside the cavity is picked up by a strongly coupled antenna. We used a coaxial copper antenna for that purpose. In addition, we used an additional weekly coupled antenna to extract the properties of the cavity by measuring the transmission coefficient of the cavity.

\begin{figure}[t]
\centerline{\includegraphics[width=.45\textwidth]{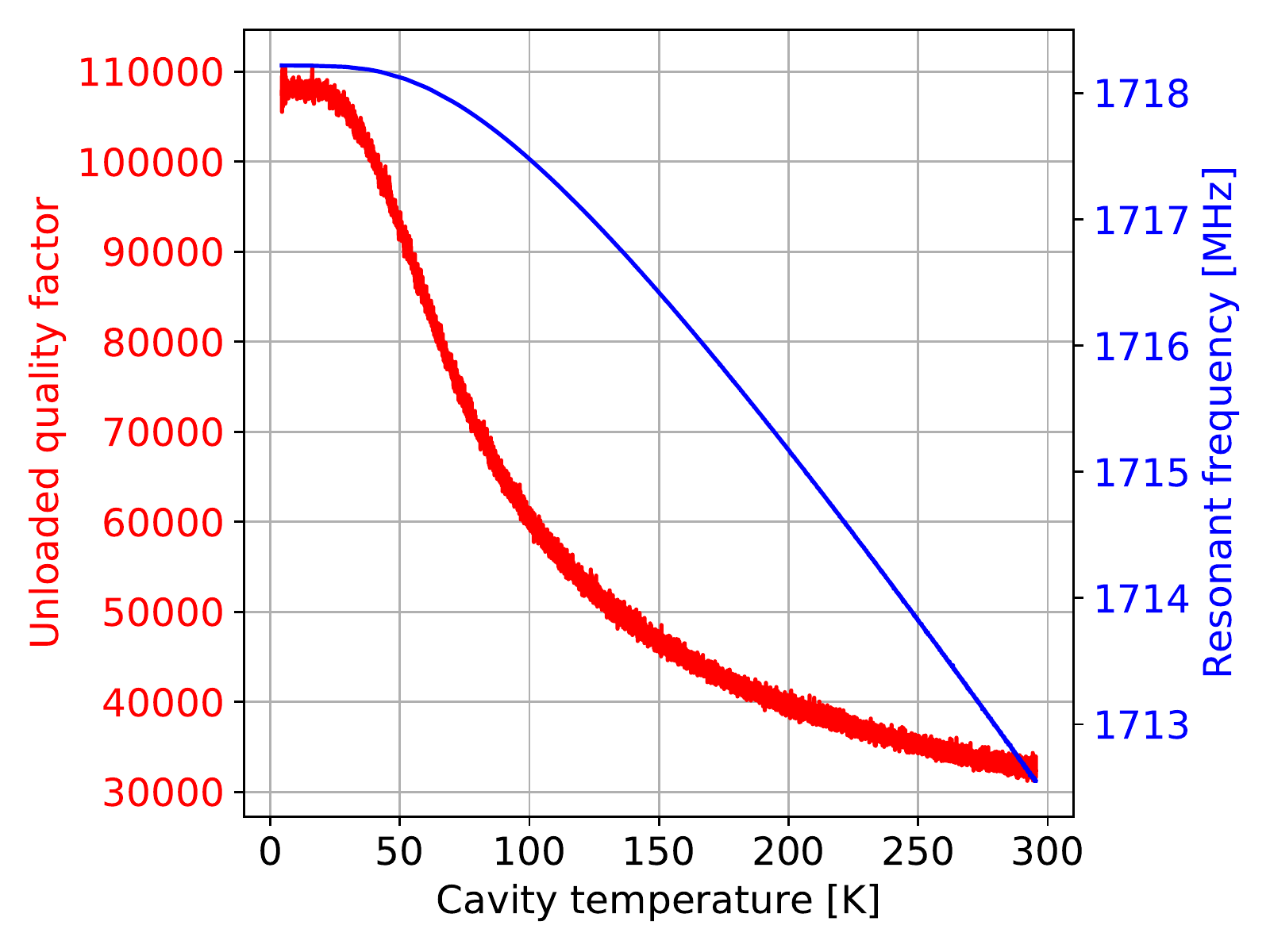}}
\caption{Unloaded quality factor (red) and resonant frequency (blue) of the TM$_{010}$ mode of the empty cavity as functions of its physical temperature. The anomalous skin effect appeared near the temperature of 30\,K.}\label{fig:cavity_q}
\end{figure}

At room temperature, the resonant frequency and unloaded quality factor ($Q_{0}$) of the TM$_{010}$ mode of the cavity were measured to be 1712.55\,MHz and 32,000, respectively. The resonant frequency was consistent with the expectation, 1712.56\,MHz, which was calculated using a resonant frequency of the TM$_{nlm}$ mode
\begin{eqnarray*}
\omega_{nlm} & = & \frac{1}{\sqrt{\mu\epsilon}}\sqrt{\frac{x^{2}_{nl}}{R^{2}}+\frac{m^{2}\pi^{2}}{d^{2}}}
\end{eqnarray*}
where $\mu$ and $\epsilon$ are the permeability and permittivity of the cavity filling, $R$ and $d$ are the inner diameter and height of the cavity, and $x_{nl}$ is $n$-th root of the $l$-th order Bessel function. It was also confirmed with simulations \cite{CST,COMSOL}. At a cryogenic temperature of 4\,K, we measured the resonant frequency and unloaded quality factor of 1718.22\,MHz and 108,000, respectively. An anomalous skin effect in the cavity is clearly shown in Figure \ref{fig:cavity_q}, as the physical temperature of the cavity reaches to a low temperature.

\section{Tuning Mechanism}
\label{sect:tuning}
In the experiment, we tuned both the resonant frequency of the TM$_{010}$ mode and the coupling coefficient of the antenna at every step.

\begin{figure}[t]
\centerline{\includegraphics[width=.45\textwidth]{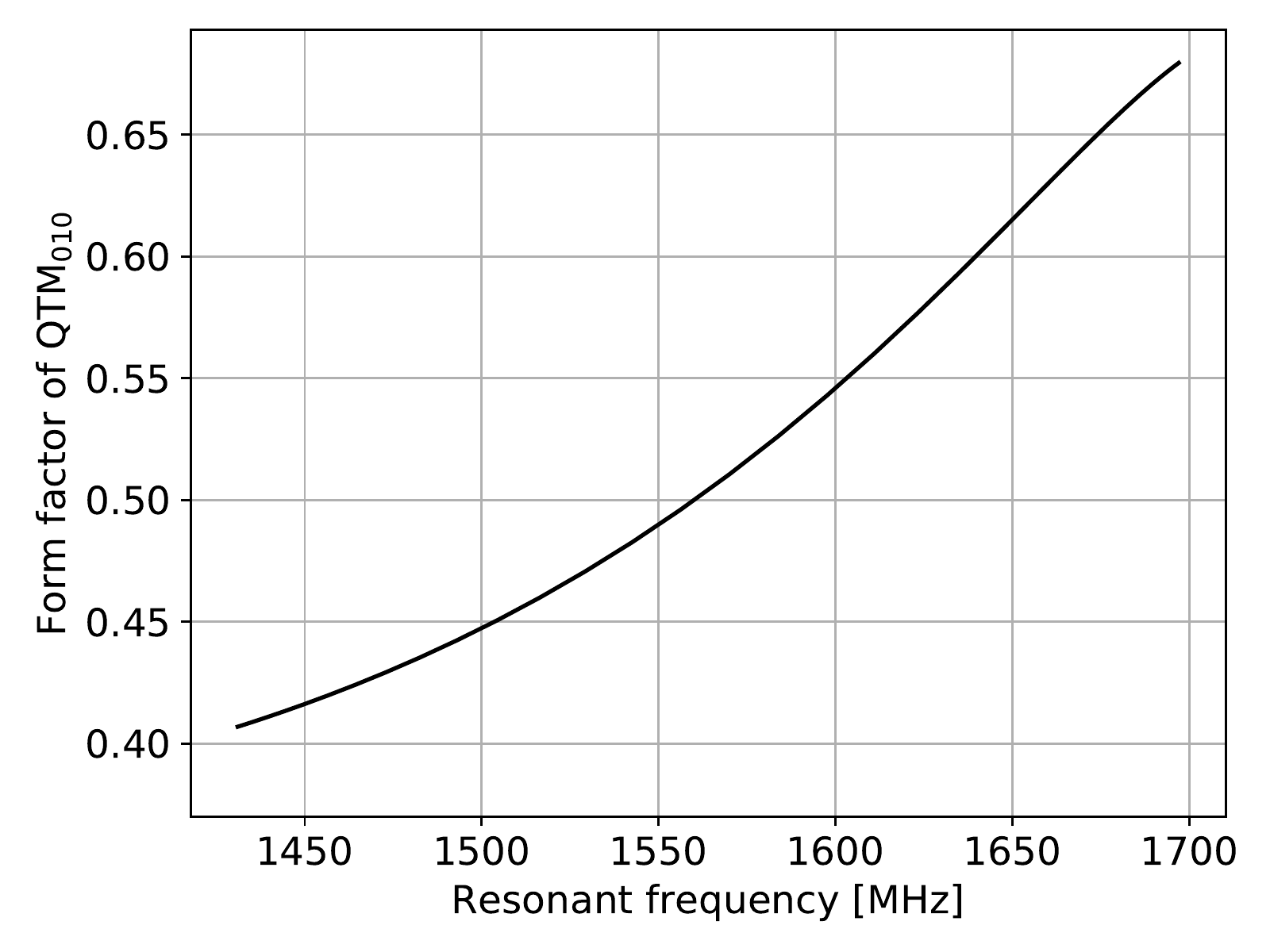}}
\caption{The form factor of the TM$_{010}$ mode obtained from simulations as a function of the resonant frequency of the cavity.}\label{fig:form_factor}
\end{figure}

The resonant frequency of the cavity was tuned using a dielectric tuning rod. The tuning rod was made of high purity Al$_{2}$O$_{3}$, also known as alumina. By rotating the tuning rod inside the cavity, the resonant frequency could be tuned from 1431 to 1697\,MHz. The form factor of the TM$_{010}$ mode, defined as
\begin{eqnarray*}
C_{010} & = & \frac{|\int d^{3}x~\mathbf{B}\cdot \mathbf{E_{010}}(\mathbf{x})|^{2}}{B_{\mathrm{avg}}^{2}V\int d^{3}x\epsilon(\mathbf{x})|\mathbf{E_{010}}(\mathbf{x})|^{2}},
\end{eqnarray*}
where $\mathbf{B}$ is the magnetic field, $\mathbf{E_{010}}$ is the electric field of the mode and $\epsilon(\mathbf{x})$ is the permittivity in the resonant cavity, is obtained from simulation studies \cite{CST,COMSOL} as shown in Figure \ref{fig:form_factor}.

\begin{figure}[t]
\centerline{\includegraphics[width=.45\textwidth]{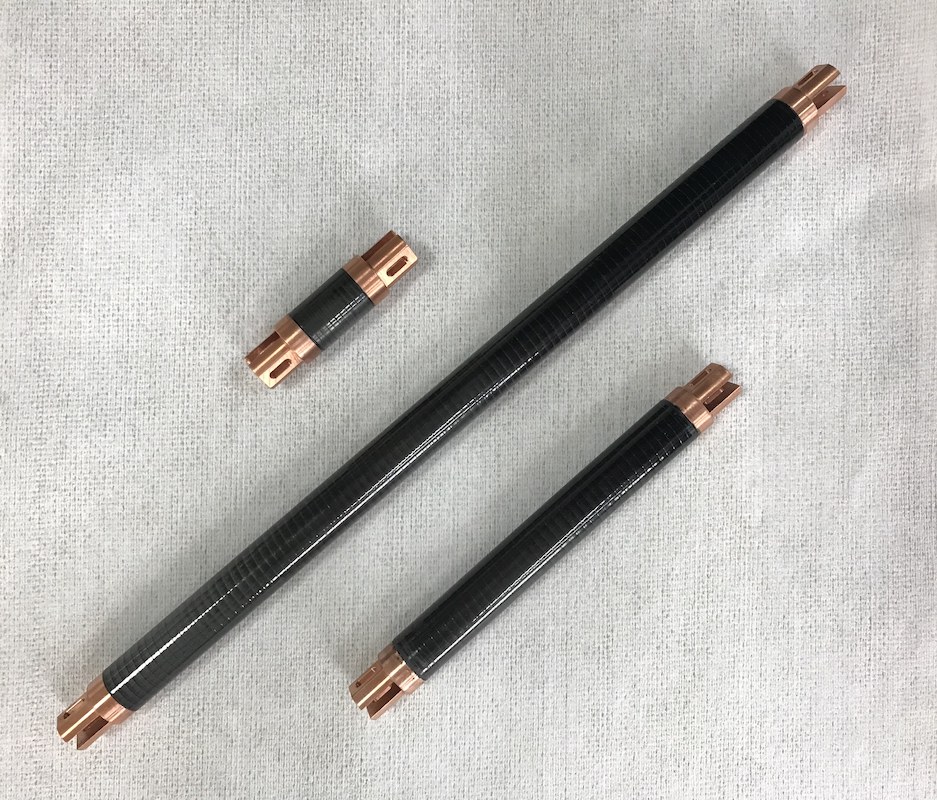}}
\caption{The driving shafts made of CFRP for the tuning mechanism.}\label{fig:cfrp_tubes}
\end{figure}

The frequency tuning system was driven by a rotational stepper motor made by Oriental motor \cite{Orientalmotor} and had a step angle of 0.0072$^{\circ}$. Since the stepper motor operates at room temperature, it was located outside the refrigerator. The motor was connected to a series of rotational drive shafts through a feed-through between the refrigerator and the stepper motor. There were six driving shafts, and each linked different temperature stages, i.e., 300\,K to 50\,K, 50\,K to 4\,K, 4\,K to 1\,K, 1\,K to 100\,mK, 100\,mK to the base temperature stage, and the base temperature stage to the resonant cavity. At each temperature stage, a thin copper stick connected the adjacent shafts, and the stick was held by two cryogenic bearings to make the whole rotational axle radially and vertically stable with minimum friction. The main body of the driving shafts is made of carbon fiber reinforced polymer (CFRP) as shown in Figure \ref{fig:cfrp_tubes} to block heat penetrations from one stage to another, and both ends of the shaft are made of copper. The shafts were also thermally linked to each temperature stage by a copper braid.

\begin{figure}[t]
\centerline{\includegraphics[width=.4\textwidth]{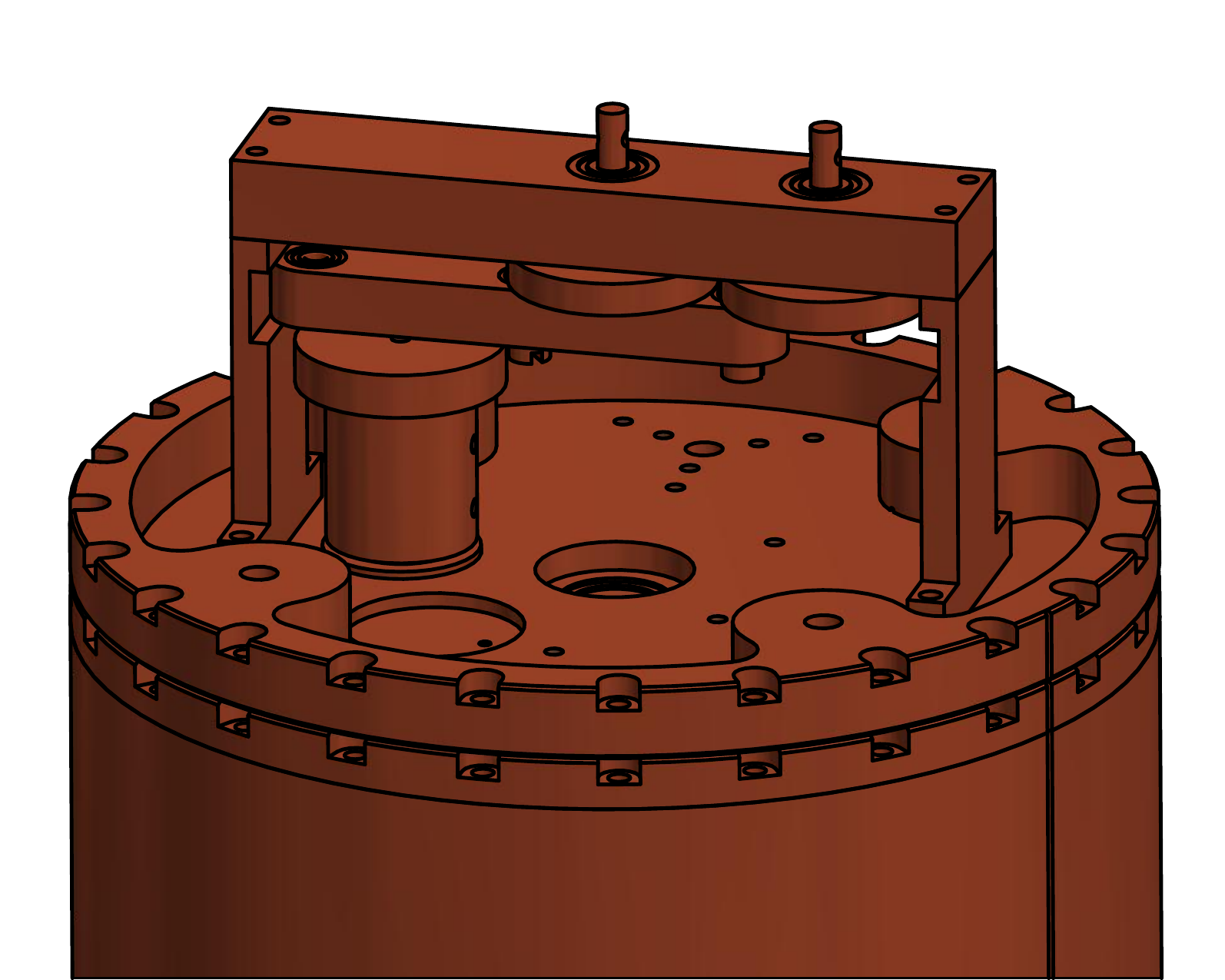}}
\caption{Locomotive frequency tuning structure. The round part in the middle is the axle of the rotational axis from the stepper motor, and the rotation is translated to the fixture on the left coupled with the tuning rod. There is an extra round fixture on the right to maintain consistent translation.}\label{fig:locomitive_tuning}
\end{figure}

Since the axle of the tuning rod of the resonant cavity is slightly offset from the axle of the rotational driving shafts, we employed a locomotive structure to translate the rotational motion from the shaft to the tuning rod as shown in Figure \ref{fig:locomitive_tuning}. To minimize friction, and therefore minimize heat generation, we used eight cryogenic bearings in the structure.

\begin{figure}[t]
\centering
\includegraphics[width=.55\textwidth]{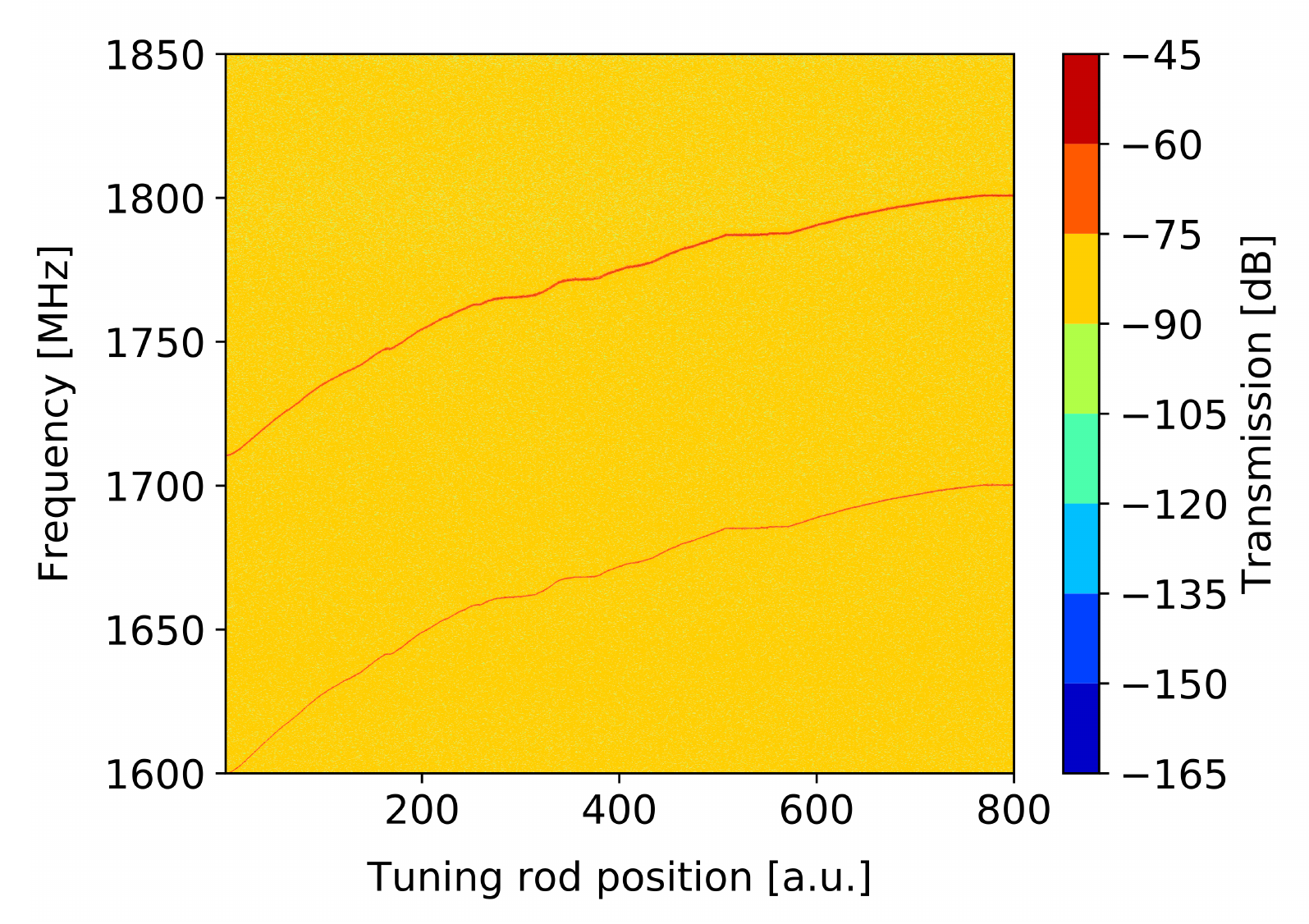}
\caption{Frequency mode map of the resonant cavity with the frequency tuning mechanism. The horizontal axis is the position in the coordinates of the tuning rod rotation. Within the frequency range of 1.60 to 1.85\,GHz, only TM$_{010}$ and TM$_{020}$ modes were found, without any mode crossing.}\label{fig:frequency_modemap}
\end{figure}

In the microwave cavity, the potential exists that the target mode, i.e., the TM$_{010}$ mode, has the same resonant frequency as the TE or TEM modes. This is so-called mode crossing, and it causes a loss of sensitivity. We confirmed that there was no mode crossing by measuring cavity transmissions as shown in Figure \ref{fig:frequency_modemap}. Accordingly, the experiment was able to search the frequency range without loss of sensitivity.

\begin{figure}[t]
\centerline{\includegraphics[width=.45\textwidth]{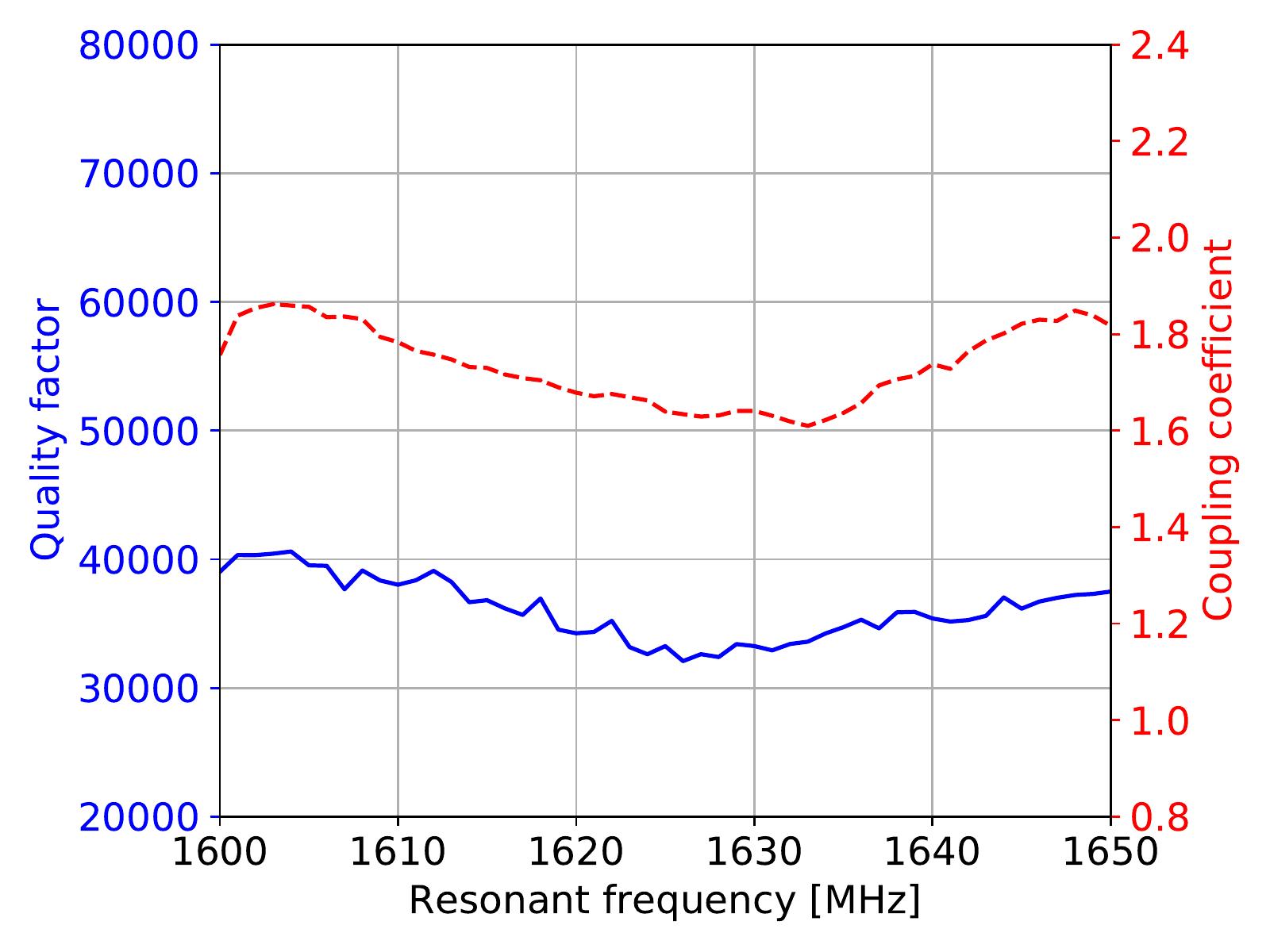}}
\caption{Measured loaded quality factor (blue solid) and antenna coupling coefficient (red dashed) of TM$_{010}$ mode of the resonant cavity as a function of resonant frequency. The coupling coefficient of the antenna varies from 1.6 to 1.9 in this measurement.}\label{fig:quality_factor}
\end{figure}

The strongly coupled antenna of the cavity was also tuned to a desired antenna coupling coefficient, $\beta$. The loaded quality factor of the cavity is inversely proportional to the coupling coefficient,
\begin{eqnarray}
Q_{L} & = & \frac{Q_{0}}{1 + \beta}.
\end{eqnarray}
The measured quality factor of the TM$_{010}$ mode as a function of resonant frequency is shown in Figure \ref{fig:quality_factor}. In the experiment, we tuned the coupling coefficient to be $1.9\pm0.1$.

\begin{figure}[t]
\centerline{\includegraphics[width=.4\textwidth]{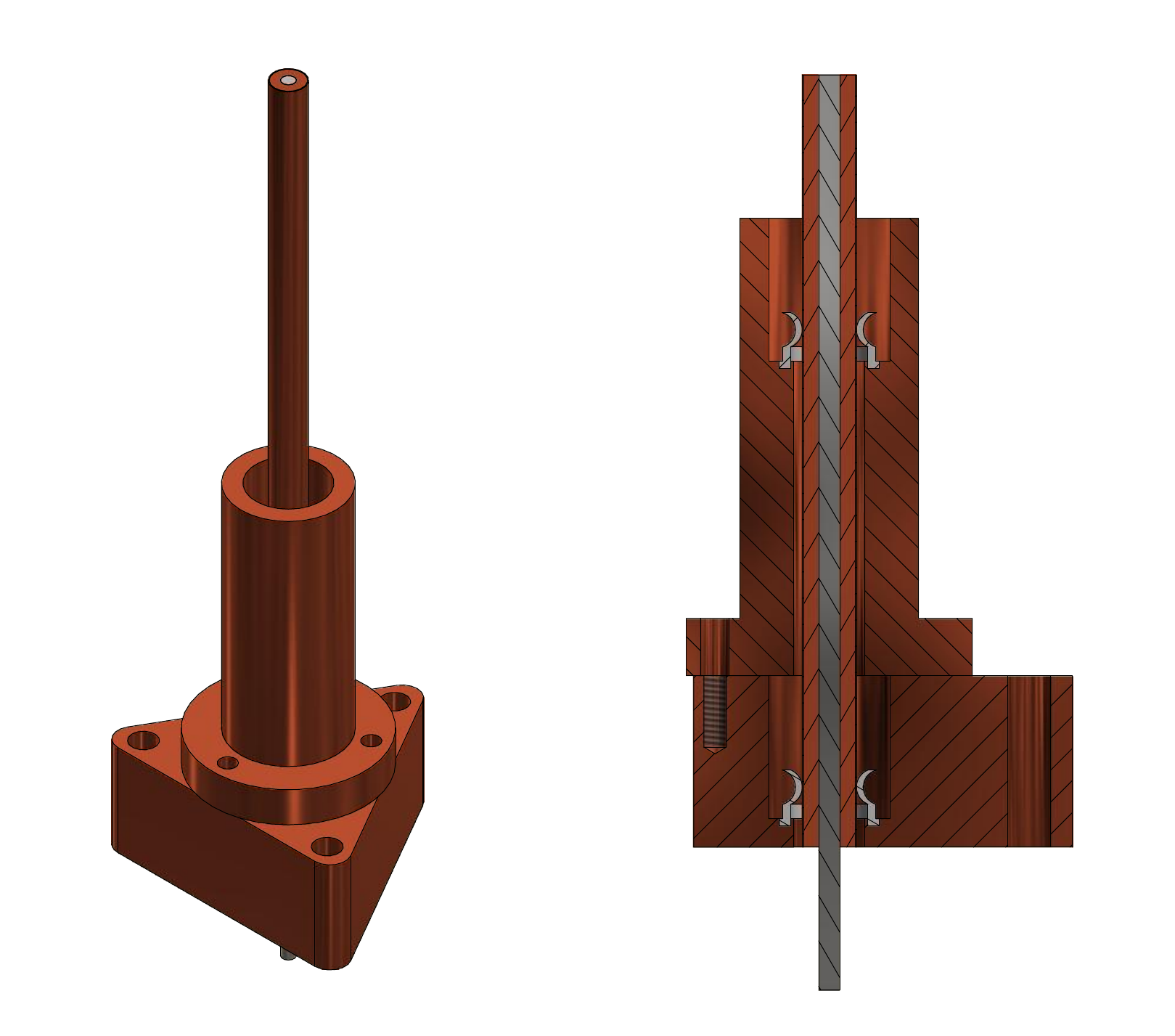}}
\caption{Antenna holder (left) and its cross-sectional view (right) with a proper ground. The surface of the coaxial antenna is held by two finger stocks on top and bottom to provide an isotropic ground.}\label{fig:antenna_grounding}
\end{figure}

The tuning system for the antenna coupling coefficient also uses a series of drive shafts similar to the frequency tuning system described above. The shafts are moved by a rotational stepper motor made by the same company. The motor was connected to the shaft through a feed-through which converted a rotational motion to a linear motion. The shafts were made of the same material as the frequency tuning system. Like the frequency tuning system, the shafts were thermally linked at each temperature stage.

To make a solid electric ground for the strongly coupled antenna, we designed a housing for the antenna as shown in Figure \ref{fig:antenna_grounding}. Since there is friction when the antenna is moved up and down, the temperature in the resonant cavity goes slightly up when the antenna coupling tuning is performed, however, the temperature quickly goes down after the tuning is done, therefore, it does not introduce a significant effect during the experiment.

The frequency and antenna coupling tunings were managed using data acquisition software, CULDAQ \cite{JPhysConfSer_898_032035_2017}. The target resonant frequency of the cavity and the desired coupling coefficient of the antenna were found by using the same algorithm. In the algorithm, the motor is rotated by an initial step size ($h_{0}$), and the software compares the current position (resonant frequency or $\beta$) to the target position. If the target position is still away in the same direction, the software rotates the motor with the same step size $h_{0}$. If the current position passes the target, the step size is reduced by a reduction factor $\alpha$, and the motor rotates with the new step size in the opposite direction, i. e., $h_{1}=h_{0}/\alpha$. This process iterates until the current position is within a given tolerance. In the experiment, we set the tolerances to be $\pm500$\,Hz and $\pm0.1$ for the resonant frequency and coupling tuning, respectively.

\begin{figure*}[t]
\centering
\includegraphics[width=.45\textwidth]{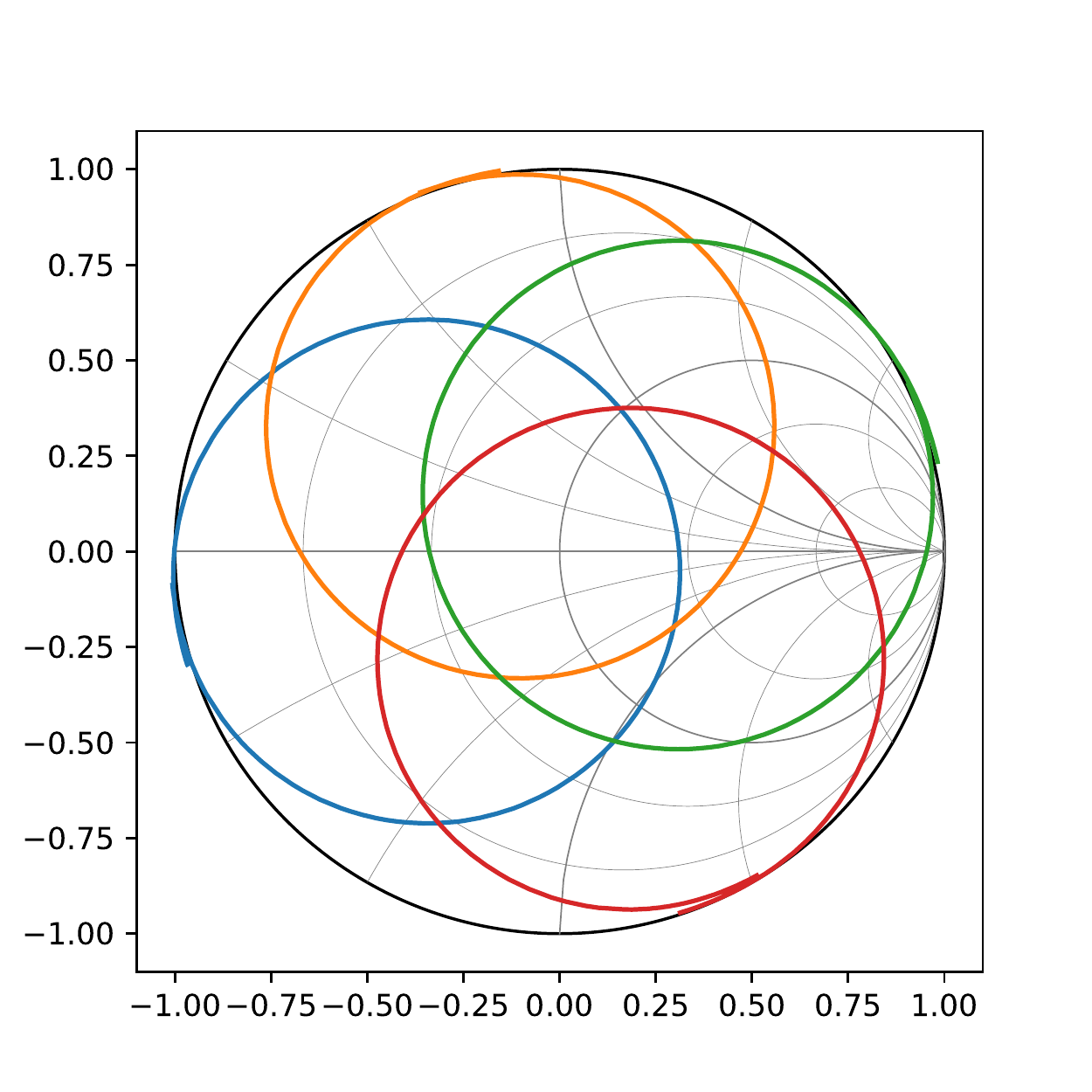}
\includegraphics[width=.45\textwidth]{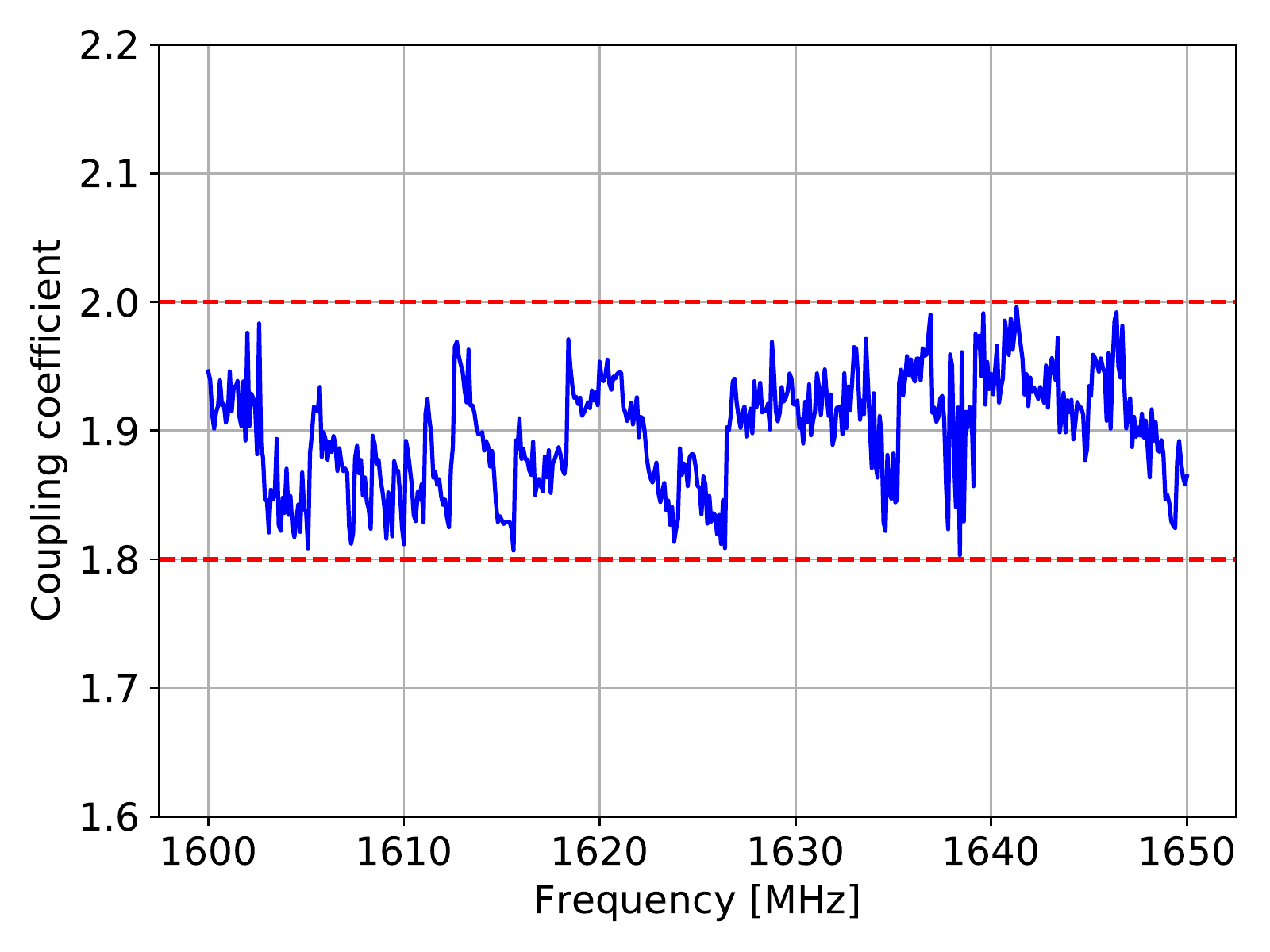}
\caption{Left: Calibrated Smith circles near the resonant frequencies of 1628\,MHz (green), 1632\,MHz (orange), 1636\,MHz (blue), and 1640\,MHz (red) with coupling coefficients of 1.90, 1.93, 1.96, and 1.94, respectively. Right: Coupling coefficients obtained over frequency with tuning. Red dashed lines show the targeted coupling coefficient range of $1.9\pm0.1$.}
\label{fig:calibration}
\end{figure*}

The antenna coupling coefficient is obtained by measuring the reflection coefficient of the strongly coupled antenna using a vector network analyzer. The Smith chart provides access to the coupling coefficient by calculating the diameter of the Smith circle \cite{IEEETransMicroTheTech_MTT32_666_1984},
\begin{eqnarray}
\beta & = & \frac{d}{2-d}
\end{eqnarray}
where $d$ is the diameter of the Smith circle. The diameter of the Smith circle is obtained by the least-square method selected data points within $-2\Delta\nu_{c}<\nu_{c}<+2\Delta\nu_{c}$, where $\nu_{c}$ and $\Delta\nu_{c}$ are the resonant frequency and the bandwidth of the cavity, respectively. If the Smith circle is not at a detuned short position, it does not provide a correct antenna coupling coefficient. To put the Smith circle at a detuned short position to obtain a correct antenna coupling coefficient, we used a loss correction function of the network analyzer with the same finding algorithm. Figure \ref{fig:calibration} shows the corrected Smith circles and tuned coupling coefficient.

Since the frequency tuning slightly changes the coupling coefficient and vice versa, we tuned the frequency and coupling together iteratively. First we performed a loss correction to put the Smith circle at a detuned short position, and the frequency tuning was performed until the resonant frequency reaches within a tolerance. Then, we tuned the antenna coupling to the desired coupling coefficient with a tolerance. This slightly changed the resonant frequency, therefore, we tune the resonant frequency again if it is out of the tolerance. If the frequency tuning takes the coupling outside the tolerance, the coupling tuning is performed again. Those tunings were performed until the loss correction, resonant frequency, and coupling coefficient were all within their tolerances. The tuning performance highly depends on the system's stability, and it typically takes around a minute for each frequency step.

\section{Microwave Receiver Chain}
\label{sec:receiver}

\begin{figure*}[t]
\centerline{\includegraphics[width=.95\textwidth]{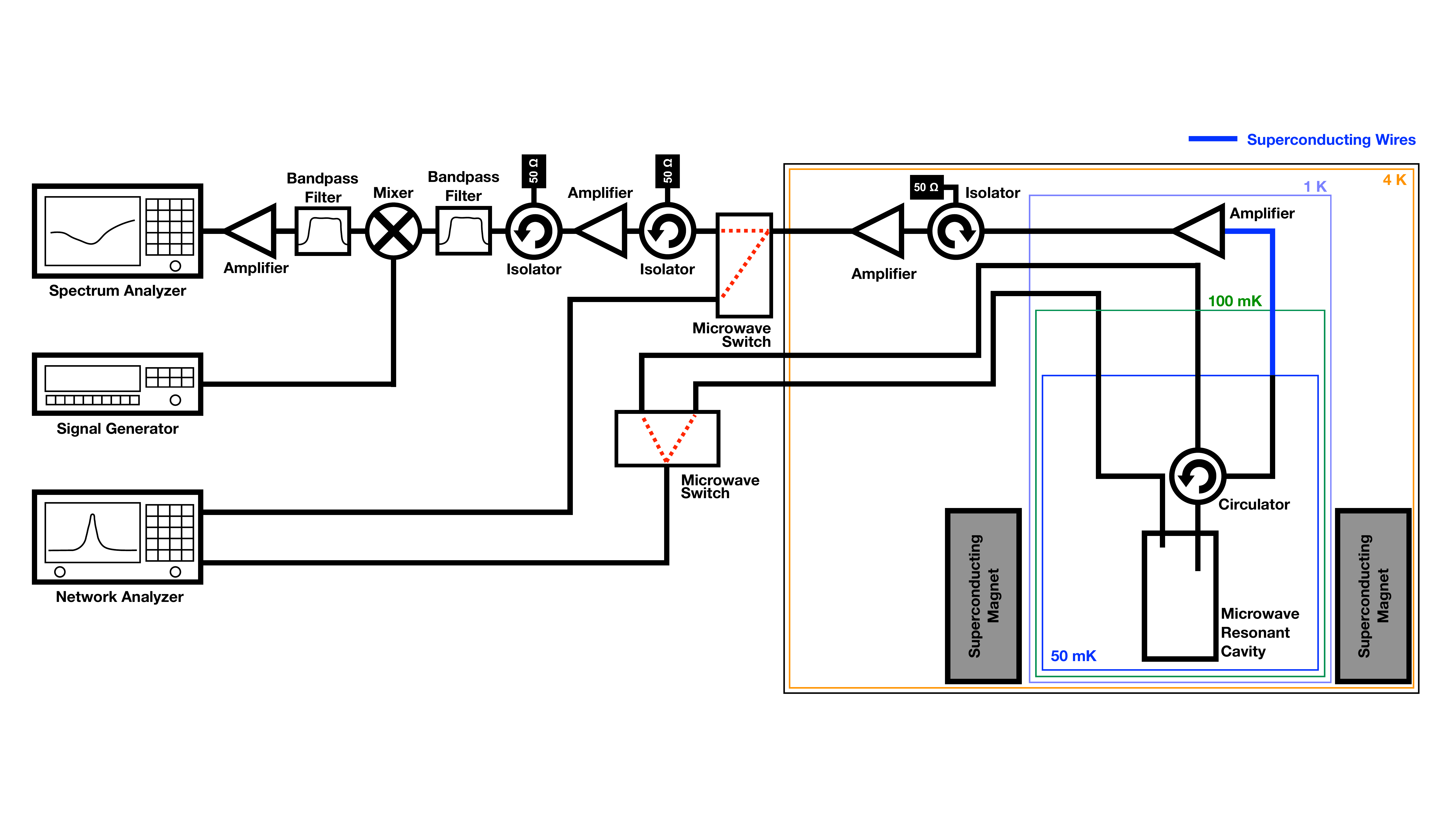}}
\caption{Microwave receiver chain of the CAPP-8TB experiment.}\label{fig:receiver_chain}
\end{figure*}

The energy stored inside the cavity is picked up by the strongly coupled antenna and acquired through a microwave receiver chain. The microwave receiver chain of the CAPP-8TB experiment is shown in Figure \ref{fig:receiver_chain}. 

To avoid reflected power from the first amplifier the first microwave component in the receiver chain is a cryogenic circulator \cite{RADITEK}. The circulator is mounted on the base temperature stage, and it covers a frequency range between 1350 and 1750\,MHz. 

The first amplifier to bring the microwaves to a detectable level is located on the 1\,K stage to minimize the physical temperature of the amplifier. We employed a cryogenic amplifier based on a high electron mobility transistor (HEMT), LNF-LNC0.6\_2A made by Low Noise Factory \cite{LowNoiseFactory}, and it covered a frequency range of 0.6 to 2\,GHz with a typical gain of 32\,dB and noise temperature of less than 1\,K. In the experiment, we maintained the physical temperature of the amplifier at 1.2\,K to maintain a stable performance. The amplifier, the 100\,mK stage, and the base temperature stage were connected by NbTi superconducting wires to reduce attenuation in the microwave path.

We used another cryogenic amplifier, LNF-LNC1\_12A \cite{LowNoiseFactory} made by the same company to further amplify the microwave power. The second amplifier, located on the 4\,K stage, covered a broader frequency range from 1 to 12\,GHz, and its typical gain and noise temperature were about 45\,dB and 5.5\,K, respectively. Since the gain of the first amplifier was high enough, the noise temperatures of the later components were negligible. We placed a cryogenic isolator \cite{RADITEK} in front of the second amplifier to eliminate the reflection from the second amplifier as well.

At room temperature, two additional amplifiers were employed to amplify the power. The first amplifier, ZRL-2400LN+ \cite{MiniCircuits}, at room temperature covered 1.0 to 2.4\,GHz with a typical gain of 31\,dB. We place isolators before and after the first amplifier to block reflections at room temperature as well. After the isolator, there was a bandpass filter, 8FV50-1550/T120 \cite{KLMicrowave}, to discard signal power that was out of the frequency region of interest. The bandpass filter only passed power within the frequency range of 1.6 to 1.7\,GHz. After the bandpass filter, we down-converted the signal to intermediate frequencies centered on 70\,MHz by using a mixer, ZX05-C24MH+ \cite{MiniCircuits}. The mixer operated in a frequency range of 0.3 to 2.4\,GHz with a typical attenuation of 5.21\,dB. We injected an RF power of 13\,dBm with a signal generator \cite{RohdeSchwarz} as a local oscillator for the mixer. The down-converted signal was cut again with another bandpass filter, SHP-50+ \cite{MiniCircuits}, which covered a frequency range of 41 to 800\,MHz. The signal was finally amplified by another amplifier, ZX60-100VH+ \cite{MiniCircuits}, which covered 0.3 to 100\,MHz with a typical gain of 37\,dB.

\begin{figure}[t]
\centerline{\includegraphics[width=.45\textwidth]{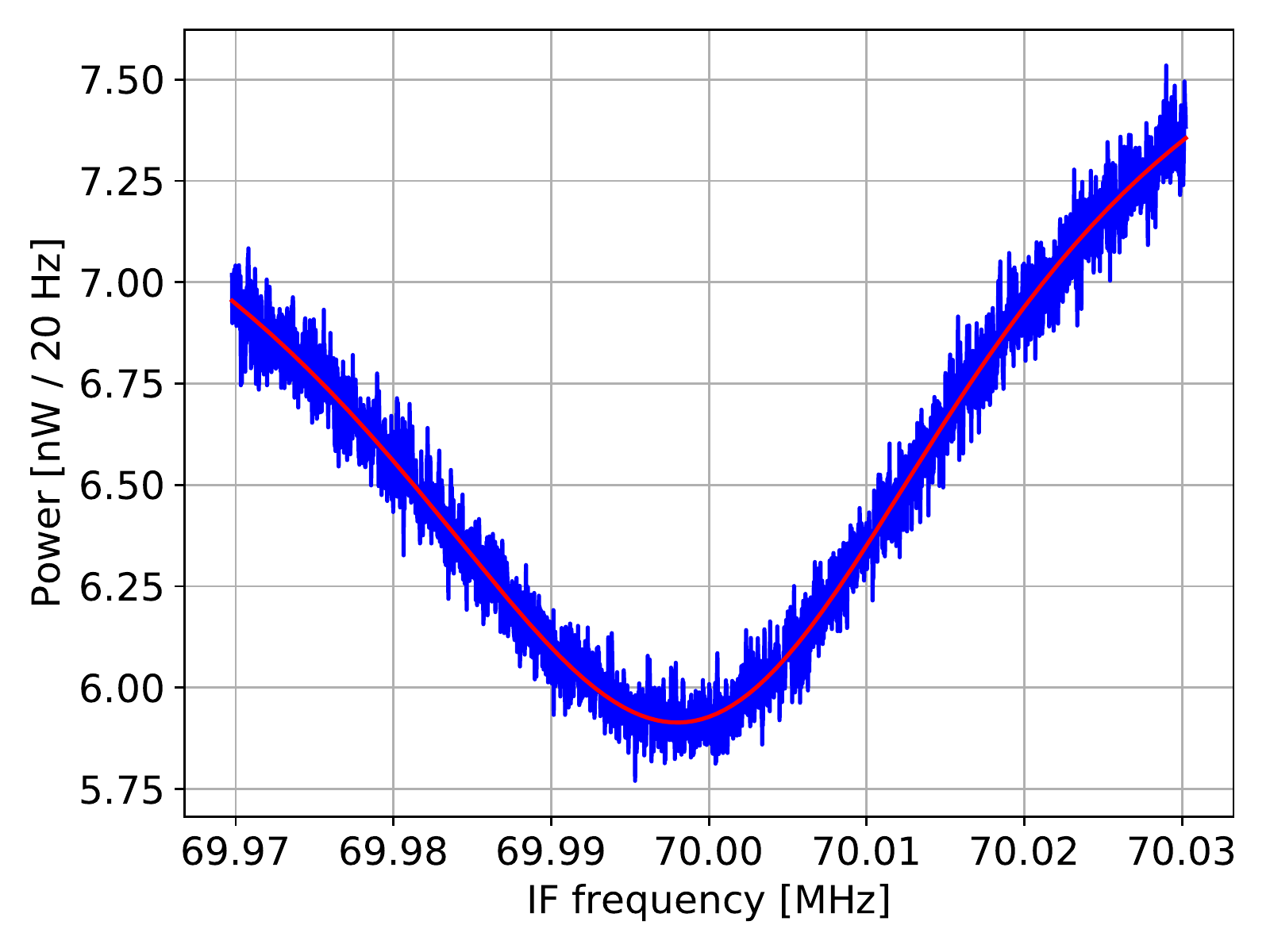}}
\caption{Typical power spectrum (blue) taken from the resonant cavity through the receiver chain. Parameterization with Eq. (\ref{eq:fiveparameterfunction}) is also shown (red) with the total system gain of about 132\,dB.}\label{fig:spectrum_sample}
\end{figure}

The power spectrum from the resonant cavity through the receiver chain was modeled using an equivalent circuit model invented by ADMX \cite{PhysRevD64_092003_2001}, 
\begin{eqnarray}
\label{eq:fiveparameterfunction}
P(\Delta) & = & k_{B}\Delta fG\left(\frac{a_{1}+8a_{3}\left(\frac{\Delta-a_{5}}{a_{2}}\right)^{2}+4a_{4}\left(\frac{\Delta-a_{5}}{a_{2}}\right)}{1+4\left(\frac{\Delta-a_{5}}{a_{2}}\right)^{2}}\right),
\end{eqnarray}
where $\Delta$ is the frequency offsets from the center of the spectrum, $k_{B}$ is the Boltzmann constant, $\Delta f$ is the resolution bandwidth, $G$ is the total system gain. The other parameters are described in Ref. \cite{PhysRevD64_092003_2001}. Figure \ref{fig:spectrum_sample} is a typical power spectrum from the resonant cavity through the receiver chain.

\begin{figure*}[t]
\centering
\includegraphics[width=.45\textwidth]{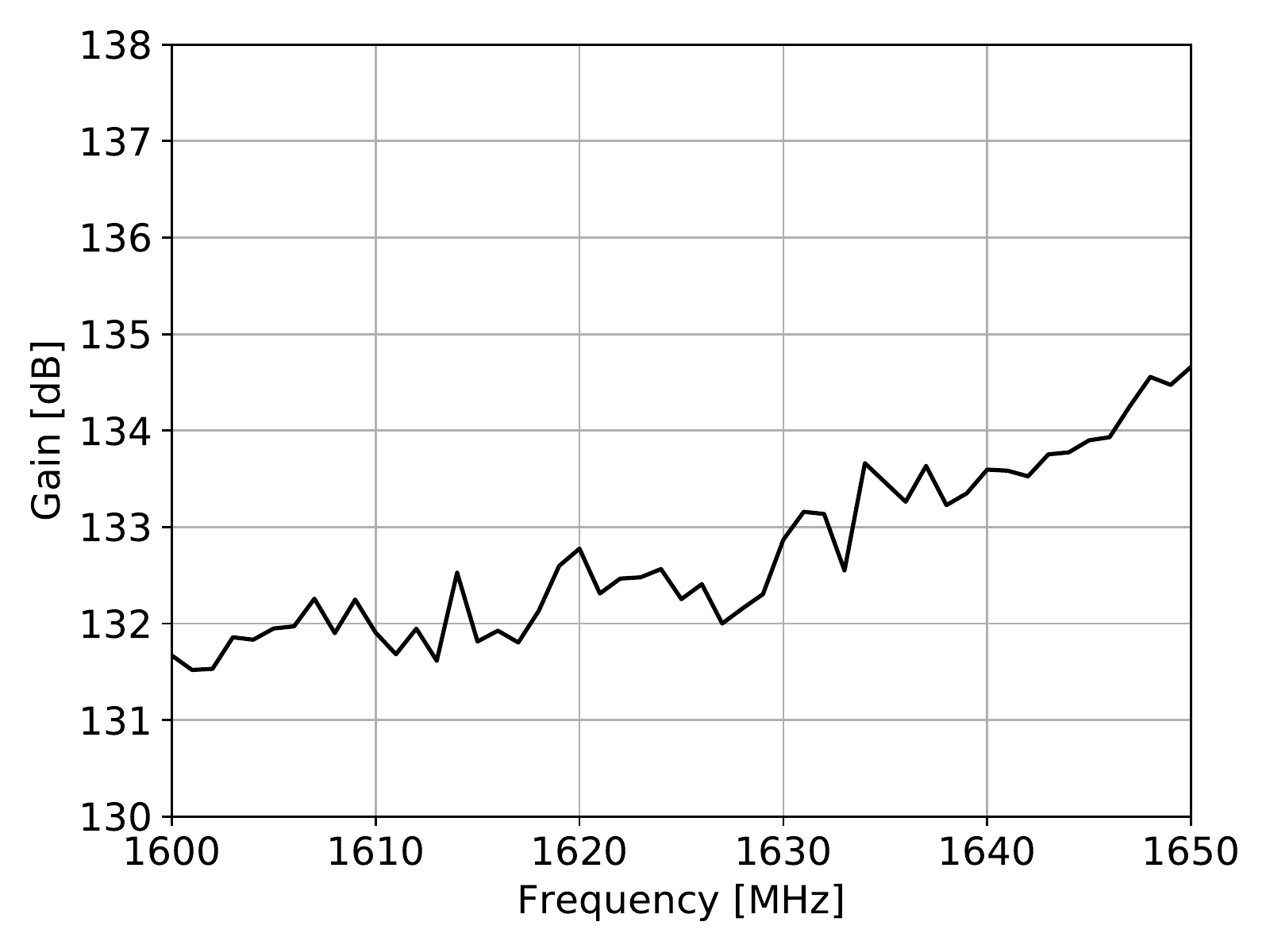}
\includegraphics[width=.45\textwidth]{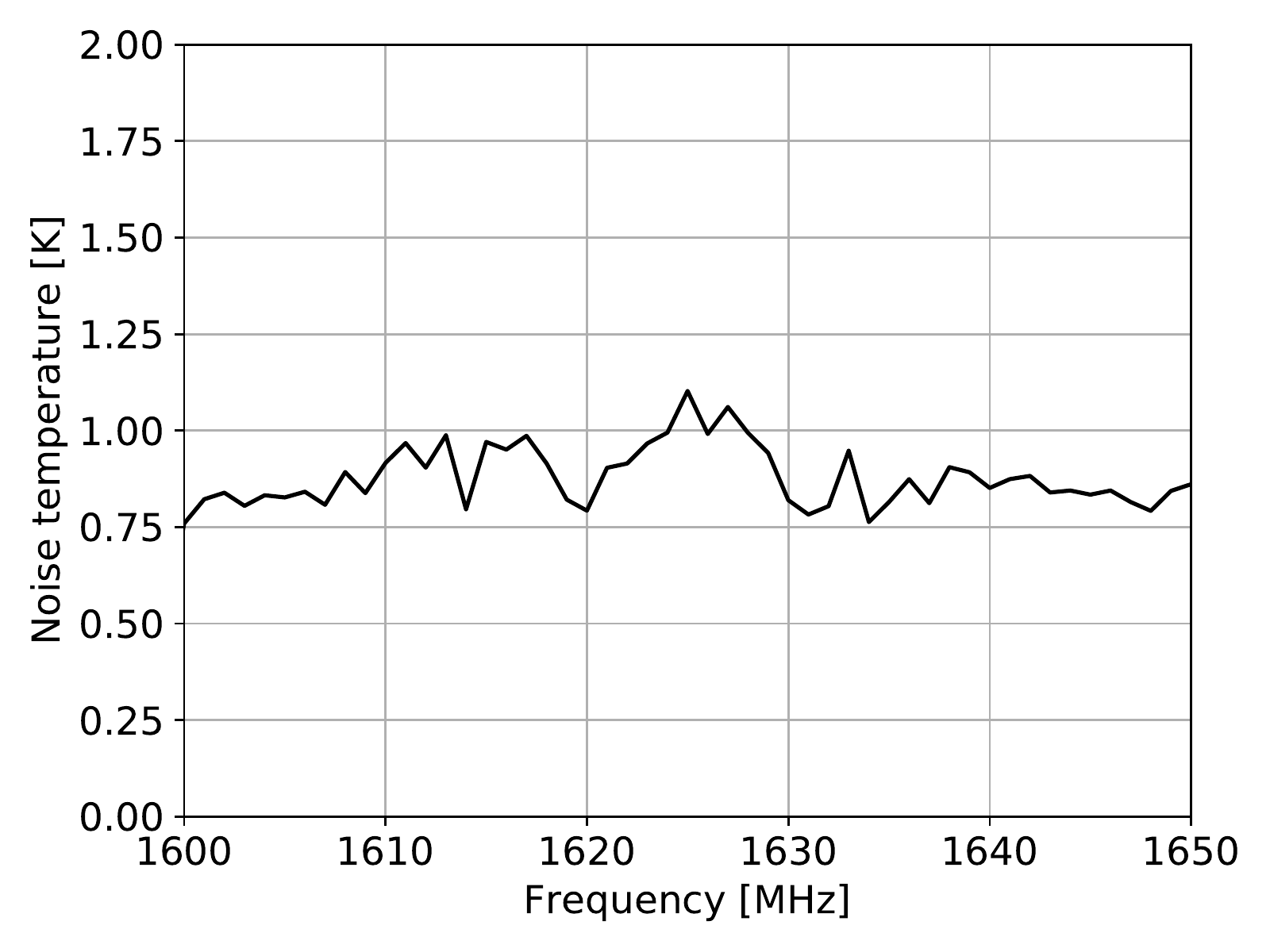}
\caption{Total system gain (left) and noise temperature (right). The uncertainty of measurements is about 7\% dominated by the systematic uncertainty from the spectrum analyzer.}
\label{fig:systemgainnoise}
\end{figure*}

The total system gain at a resonant frequency is
\begin{eqnarray}
\label{eq:gainfunction}
G & = & \frac{P_{h}-P_{c}}{k_{B}\Delta f(T_{h}-T_{c})}
\end{eqnarray}
where $P_{h}$ and $P_{c}$ are the powers measured at different cavity temperatures of $T_{h}$ and $T_{c}$, respectively. There were negligible changes in the noise temperature of the chain except for the resonant cavity and the quality factor of the resonant cavity over the temperature difference, $T_{h}-T_{c}$. We obtained the total system gain using Eq. (\ref{eq:gainfunction}) with the power measurements at the cavity temperatures of  50 and 200\,mK, as shown in Figure \ref{fig:systemgainnoise} (left). The system noise temperature including the physical temperature of the resonant cavity was parameterized using Eq. (\ref{eq:fiveparameterfunction}), and is shown in Figure \ref{fig:systemgainnoise} (right).

We also measured the cavity properties using a vector network analyzer \cite{RohdeSchwarz}. Input and output paths were connected to the network analyzer for this purpose. Measurements of the power spectra and properties of the cavity were switched using microwave switches \cite{MiniCircuits}. During the experiment, we measured cavity properties such as the loaded quality factor and coupling coefficient by taking the transmission and reflection coefficients, respectively, and those properties were used in the data analysis as auxiliary data.

\section{Data Acquisition}
The power spectrum taken from the resonant cavity through the receiver chain was acquired using a spectrum analyzer \cite{RohdeSchwarz}. Also, auxiliary data such as cavity properties, physical temperatures, and magnetic field were measured using a network analyzer, temperature controller, and magnet controller, respectively. Those devices were all connected to a data acquisition computer through various interfaces such as Ethernet, RS-232, GPIB, and USB. They were controlled and monitored using a home-grown data acquisition software, CULDAQ \cite{JPhysConfSer_898_032035_2017}. The software records the data in ROOT \cite{ROOT} format to make it easy to handle.

\begin{figure}[t]
\centering
\includegraphics[width=.45\textwidth]{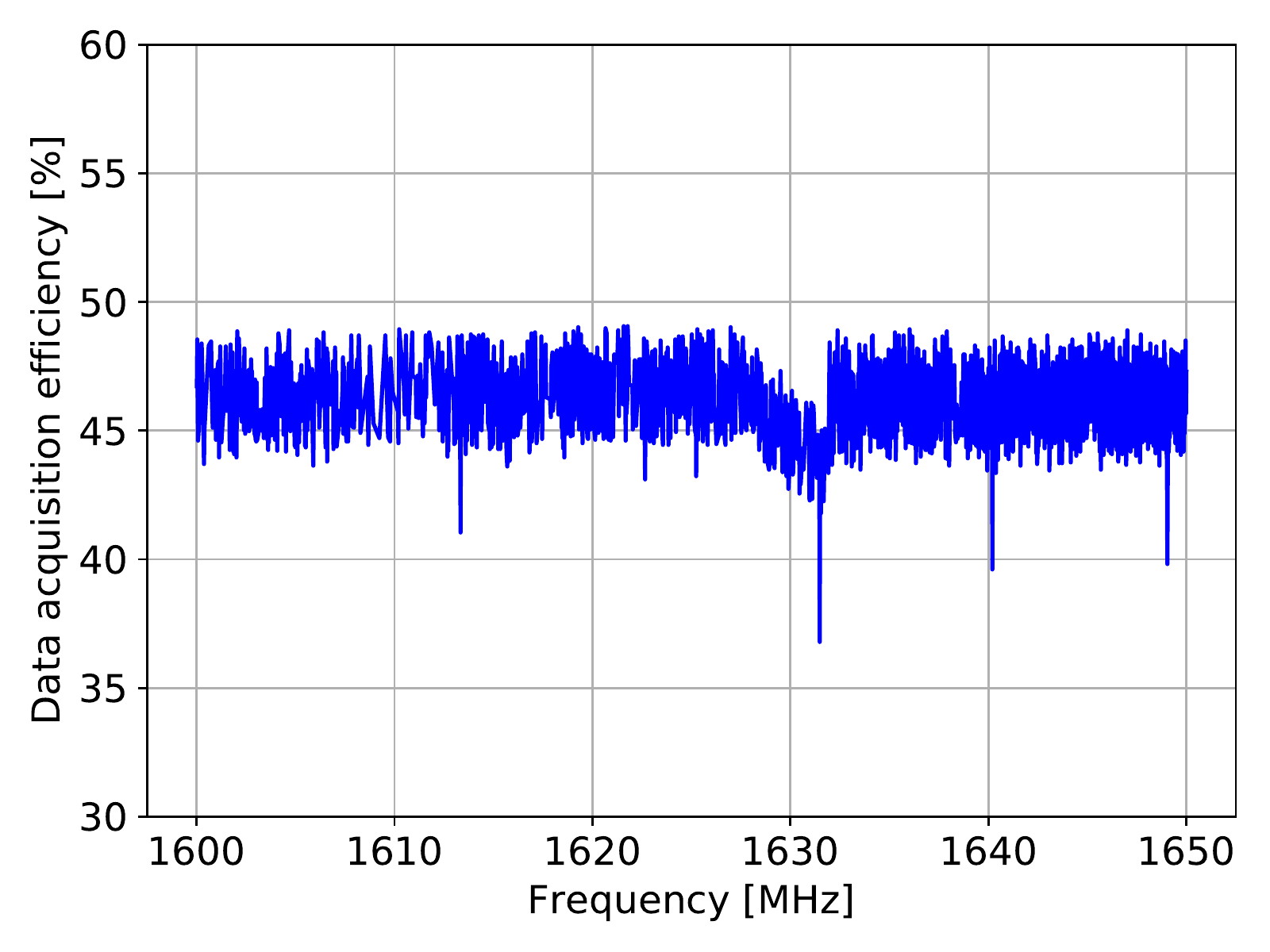}
\caption{Overall data acquisition efficiency of the experiment for 12,000 power spectra averaged at each frequency step. Efficiency drops at several frequencies are from tuning performance drops due to unstable tuning region of the system.}
\label{fig:daq_efficiency}
\end{figure}

The efficiency of data acquisition highly depends on the performance of the spectrum analyzer. When the parameters of the resolution bandwidth were 20\,Hz with a frequency span of 60.48\,kHz, the data acquisition efficiency of the spectrum analyzer was found to be about 47\%. Including the frequency and coupling tuning with calibrations, the overall data acquisition efficiency was found to be around 46\% for 12,000 power spectra, averaged at each frequency step as shown in Figure \ref{fig:daq_efficiency}.

The signal-to-noise ratio is defined as the proportion of signal power to fluctuation of noise power,
\begin{eqnarray}
\mathrm{SNR} & = & \frac{P_{S}}{\delta P_{n}},
\end{eqnarray}
and the noise fluctuation can be statistically reduced by $\sqrt{N}$ where $N$ is the number of samples taken. From the radiometer equation, the noise power is determined using a system noise temperature, $P_{n}=k_{B}b_{a}T_{n}$, therefore, the SNR can be rewritten as
\begin{eqnarray}
\mathrm{SNR} & = & \sqrt{N}\frac{P_{S}}{k_{B}b_{a}T_{n}}
\end{eqnarray}
where $b_{a}$ is the axion signal window.

In principle, the experiment could achieve the detection sensitivity of any axion model by taking samples for a long time, however, external effects such as vibration of the system or fluctuations in receiver components may affect the sensitivity. This makes it necessary to test the detection sensitivity of the experiment. 

For this test, we took 4M spectra at a resonant frequency, and subtracted the baseline with a parameterization using Eq. (\ref{eq:fiveparameterfunction}). The baseline subtracted power is expected to be centered at zero with the width of noise fluctuation. In ideal conditions, the width of the baseline subtracted power is equal to the expected noise fluctuation, $k_{B}b_{a}T_{n}/\sqrt{N}$. If the fitted width differs from the expectation for a certain number of averaged spectra, it implies that the system is not sensible to the detection sensitivity corresponding to the number of averaged spectra. 

\begin{figure}[t]
\centering
\includegraphics[width=.45\textwidth]{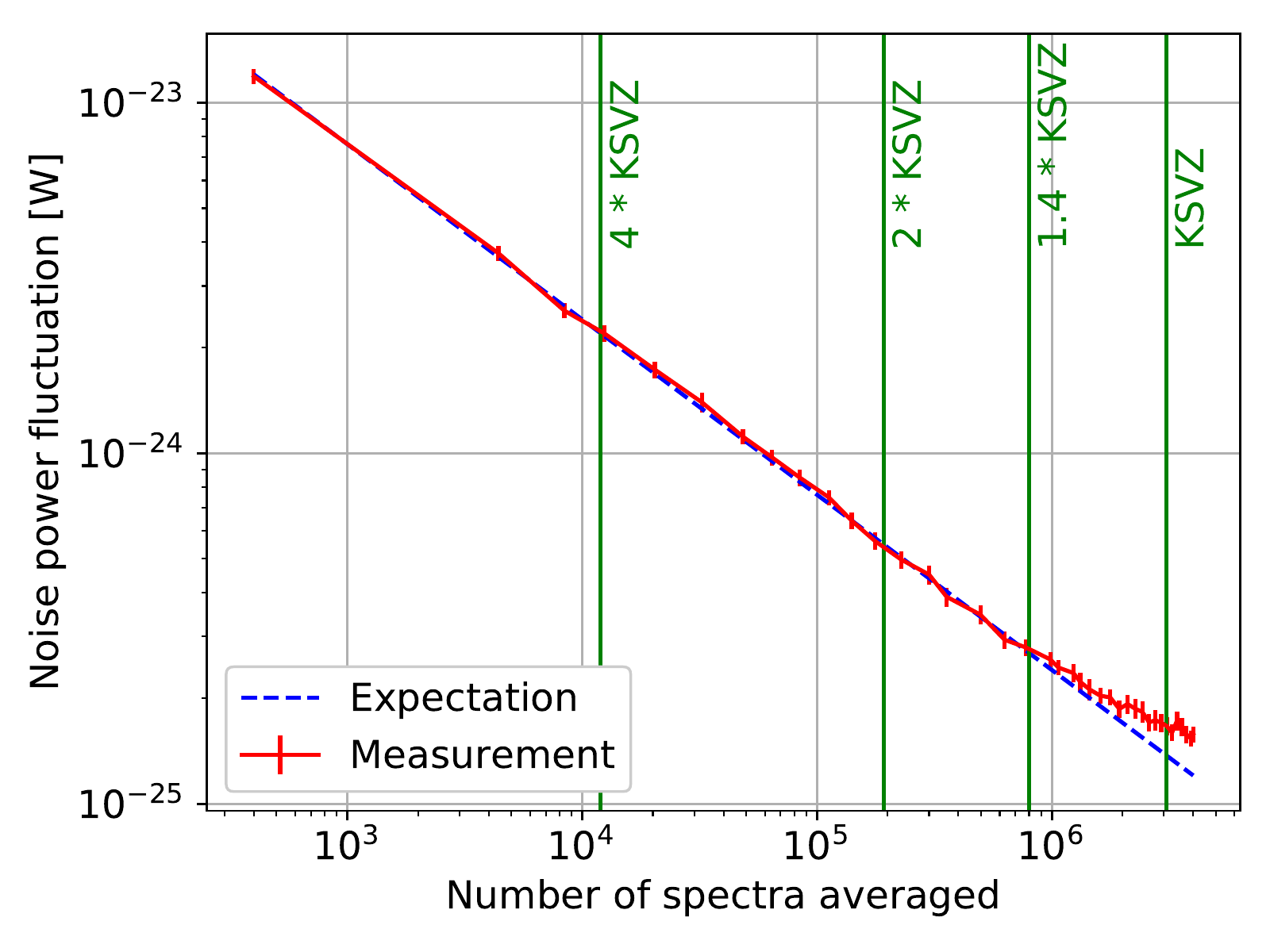}
\caption{Detection sensitivity of the CAPP-8TB experiment (red point) and the expectation (blue dashed line). The numbers of spectra required to reach levels of sensitivity are also shown in green vertical lines.}
\label{fig:detectorresolution}
\end{figure}

Figure \ref{fig:detectorresolution} shows the detection sensitivity of CAPP-8TB. The system is sensible as expected up to about 800k power spectra averaged, and it does not follow the expectation with more power spectra averaged. Therefore, we claim the system is capable of reaching $g_{\gamma}\ge1.4\times g_{\gamma}^{\mathrm{KSVZ}}$. With the current configuration, however, it would take about 7.4 years for a frequency scan of 50\,MHz to reach $g_{\gamma}=1.4\times g_{\gamma}^{\mathrm{KSVZ}}$, therefore, we set a practical goal of sensitivity $g_{\gamma}\gtrsim4\times g_{\gamma}^{\mathrm{KSVZ}}$ for the experiment.

\section{Phase 1 Physics Run}
We took data within the frequency range of 1600\,MHz to 1650\,MHz (a mass range of 6.62 to 6.82\,$\mu$eV, equivalently) with a sensitivity of $g_{\gamma}\gtrsim4\times g_{\gamma}^{\mathrm{KSVZ}}$, which is the upper boundary of the QCD axion band \cite{PhysRevD52_3132_1995}, from September 25 to November 11 in 2019 including a week of maintenance for the refrigerator.

In the run, we averaged 400 spectra over 30 times, therefore, took 12000 spectra averaged at each frequency step with a resolution bandwidth of 20\,Hz. The frequency span was set to be 60.48\,kHz, therefore, there were 3025 frequency bins in a frequency step. We tuned the resonant frequency with a frequency step size of 20\,kHz, therefore, 2501 frequency steps were scanned. At each frequency step, a measurement took about 23 minutes, including frequency and coupling tunings with calibrations.

\begin{figure}[t]
\centering
\includegraphics[width=.45\textwidth]{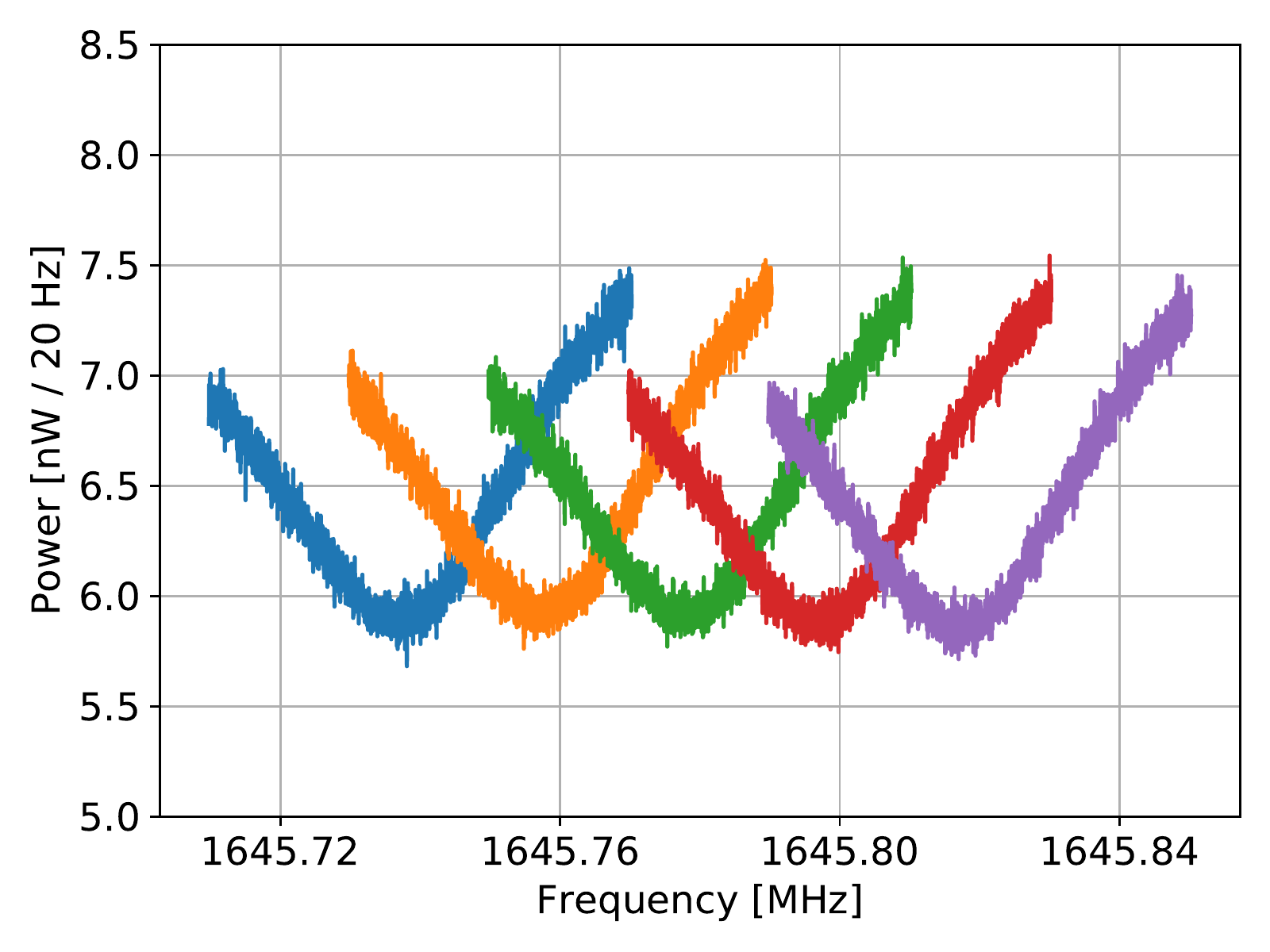}
\caption{Power spectra taken from the phase 1 physics run at several frequency steps. At each frequency step, 12,000 spectra are averaged.}
\label{fig:physicsrun_powerspectra}
\end{figure}

A subset of the power spectra taken from the experiment is shown in Figure \ref{fig:physicsrun_powerspectra}, and the experimental results can be found in Ref. \cite{PhysRevLett124_101802_2020}.

\section{Prospects}
The experiment reached a sensitivity of $g_\gamma\gtrsim4\times g_{\gamma}^{\mathrm{KSVZ}}$ within a mass range of 6.62 to 6.82\,$\mu$eV as described in the previous section. The next run is scanning other frequency ranges with the same sensitivity. Since the system is capable of scanning from 1431\,MHz to 1697\,MHz, it can be easily extended to other frequency ranges with minimum changes, such as the bandpass filter described in Section \ref{sec:receiver}. We are now scanning a mass range of 6.20 to 6.62\,$\mu$eV which corresponds to a frequency range of 1500 to 1600\,MHz as the phase 2 physics run.

As shown in Eq. (\ref{eq:scan_rate}), the scan rate of the experiment is proportional to $\eta B_{\mathrm{avg}}^{4}V^{2}C_{010}^{2}Q_{L}$ and inversely proportional to $T^{2}_{n}$. $B_{\mathrm{avg}}$ and $V$ are limited by the superconducting magnet, and $Q_{L}$ is hardly enhanced with a normal conducting material. Therefore, $\eta$ and $T_{n}$ are easy factors to improve as an upgrade to the experiment.

Since the experiment uses a commercial spectrum analyzer, there is an opportunity to increase the data acquisition efficiency by employing a fast data acquisition system. Using a digitizer, time domain data can be taken in nearly real time, and the efficiency could be more than 90\%, which is almost a factor of two improvement from the current configuration.

The system noise temperature, $T_{n}$ can also be minimized by employing a quantum noise limited amplifier such as a Josephson parametric amplifier (JPA) or a microstrip SQUID amplifier (MSA). With an ideal quantum-limited noise, it will decrease the system noise temperature by almost a factor of seven.

Those system upgrades will provide about two orders of magnitude faster scan rate, and allow us to search for axions near the KSVZ region. Both upgrades are currently being studied.

\section{Conclusions}
The CAPP-8TB experiment is an axion haloscope dedicated to searching a mass range around 6.7\,$\mu$eV with a tunable microwave resonant cavity under a magnetic field of 8\,T. We have built the experimental system, which is capable of searching $g_\gamma\ge1.4\times g_{\gamma}^{\mathrm{KSVZ}}$. The first run of the experiment has scanned a mass range of 6.62 to 6.82\,$\mu$eV with a sensitivity of $g_\gamma\gtrsim4\times g_{\gamma}^{\mathrm{KSVZ}}$, and the result is reported in Ref. \cite{PhysRevLett124_101802_2020}. The next run for a mass range of 6.20 to 6.62\,$\mu$eV with the same sensitivity is in progress.

\section*{Acknowledgements}
This work was supported by IBS-R017-D1-2020-a00.


\end{document}